\documentclass[prl,reprint,superscriptaddress,floatfix]{revtex4-1}
\usepackage{graphicx}
\usepackage{color}
\usepackage{dcolumn}
\usepackage{bm}
\usepackage{amssymb}
\usepackage{color,amsmath}
\usepackage{soul,xcolor} 
\usepackage{natbib}
\setstcolor{red}
\usepackage{epstopdf}
\usepackage[normalem]{ulem}

\begin{document}
\title{Intrinsic spin Hall torque in a moir\'e Chern magnet}
\author{C. L. Tschirhart}
\thanks{These authors contributed equally to this work}
\affiliation{Department of Physics, University of California at Santa Barbara, Santa Barbara CA 93106, USA}
\author{Evgeny Redekop}
\thanks{These authors contributed equally to this work}
\affiliation{Department of Physics, University of California at Santa Barbara, Santa Barbara CA 93106, USA}
\author{Lizhong Li}
\affiliation{School of Applied and Engineering Physics, Cornell University, Ithaca, NY, USA}
\author{Tingxin Li}
\affiliation{School of Applied and Engineering Physics, Cornell University, Ithaca, NY, USA}
\author{Shengwei Jiang}
\affiliation{School of Applied and Engineering Physics, Cornell University, Ithaca, NY, USA}
\author{T. Arp}
\affiliation{Department of Physics, University of California at Santa Barbara, Santa Barbara CA 93106, USA}
\author{O. Sheekey}
\affiliation{Department of Physics, University of California at Santa Barbara, Santa Barbara CA 93106, USA}
\author{Takashi Taniguchi}
\affiliation{International Center for Materials Nanoarchitectonics,
National Institute for Materials Science,  1-1 Namiki, Tsukuba 305-0044, Japan}
\author{Kenji Watanabe}
\affiliation{Research Center for Functional Materials,
National Institute for Materials Science, 1-1 Namiki, Tsukuba 305-0044, Japan}
\author{Kin Fai Mak}
\affiliation{School of Applied and Engineering Physics, Cornell University, Ithaca, NY, USA}
\affiliation{Laboratory of Atomic and Solid State Physics, Cornell University, Ithaca, NY, USA}
\affiliation{Kavli Institute at Cornell for Nanoscale Science, Ithaca, NY, USA}
\author{Jie Shan}
\affiliation{School of Applied and Engineering Physics, Cornell University, Ithaca, NY, USA}
\affiliation{Laboratory of Atomic and Solid State Physics, Cornell University, Ithaca, NY, USA}
\affiliation{Kavli Institute at Cornell for Nanoscale Science, Ithaca, NY, USA}
\author{A. F. Young}
\email{andrea@physics.ucsb.edu}
\affiliation{Department of Physics, University of California at Santa Barbara, Santa Barbara CA 93106, USA}

\maketitle
 
\textbf{In spin torque magnetic memories, electrically actuated spin currents are used to switch a magnetic bit.
Typically, these require a multilayer geometry including both a free ferromagnetic layer and a second layer providing spin injection.  For example, spin may be injected by a nonmagnetic layer exhibiting a large spin Hall effect\cite{hidding_spin-orbit_2020,shao_roadmap_2021}, a phenomenon known as spin-orbit torque.
Here, we demonstrate a spin-orbit torque magnetic bit in a single two-dimensional system with intrinsic magnetism and strong Berry curvature. 
We study AB-stacked MoTe$_2$/WSe$_2$, which hosts a magnetic Chern insulator at a carrier density of one hole per moir\'e superlattice site\cite{li_quantum_2021}.  
We observe hysteretic switching of the resistivity as a function of applied current. 
Magnetic imaging using a superconducting quantum interference device reveals that current switches correspond to reversals of individual magnetic domains. 
The real space pattern of domain reversals aligns precisely with spin accumulation measured near the high-Berry curvature Hubbard band edges.  
This suggests that intrinsic spin- or valley-Hall torques drive the observed current-driven magnetic switching in both MoTe$_2$/WSe$_2$ and other moir\'e materials\cite{sharpe_emergent_2019,serlin_intrinsic_2020}. 
The switching current density of $10^3 A\cdot cm^{-2}$ is significantly less than reported in other platforms\cite{fan_magnetization_2014,jiang_efficient_2019,nair_electrical_2020} paving the way for efficient control of magnetic order.}

To support a magnetic Chern insulator and thus exhibit a quantized anomalous Hall (QAH) effect, a two dimensional electron system must host both spontaneously broken time-reversal symmetry and topologically nontrivial bands\cite{chang_quantum_2022}. 
This makes Chern magnets ideal substrates upon which to engineer low-current magnetic switches, because the same Berry curvature responsible for the nontrivial band topology also produces spin- or valley-Hall effects that may be used to effect magnetic switching.  
Recently, moir\'e heterostructures emerged as a versatile platform for realizing intrinsic Chern magnets\cite{chen_tunable_2020-1,serlin_intrinsic_2020,polshyn_electrical_2020,li_quantum_2021}.  
In these systems, two layers with mismatched lattices are combined, producing a long-wavelength moir\'e pattern that reconstructs the single particle band structure within a reduced superlattice Brillouin zone. 
In certain cases, moir\'e heterostructures host superlattice minibands with narrow bandwidth, placing them in a strongly interacting regime where Coulomb repulsion may lead to one or more broken symmetries\cite{balents_superconductivity_2020,andrei_marvels_2021}.  
In several such systems, the underlying bands are topologically nontrivial\cite{zhang_nearly_2019,zhang_spin-textured_2021}, setting the stage for the appearance of anomalous Hall effects when combined with time-reversal symmetry breaking\cite{sharpe_emergent_2019}. Notably, in twisted bilayer graphene low current magnetic switching has been observed\cite{sharpe_emergent_2019,serlin_intrinsic_2020}, though consensus does not exist on the underlying mechanism\cite{he_giant_2020,su_current-induced_2020,ying_current_2021}.  

\begin{figure}[ht!]
 \includegraphics[width=3.5in]{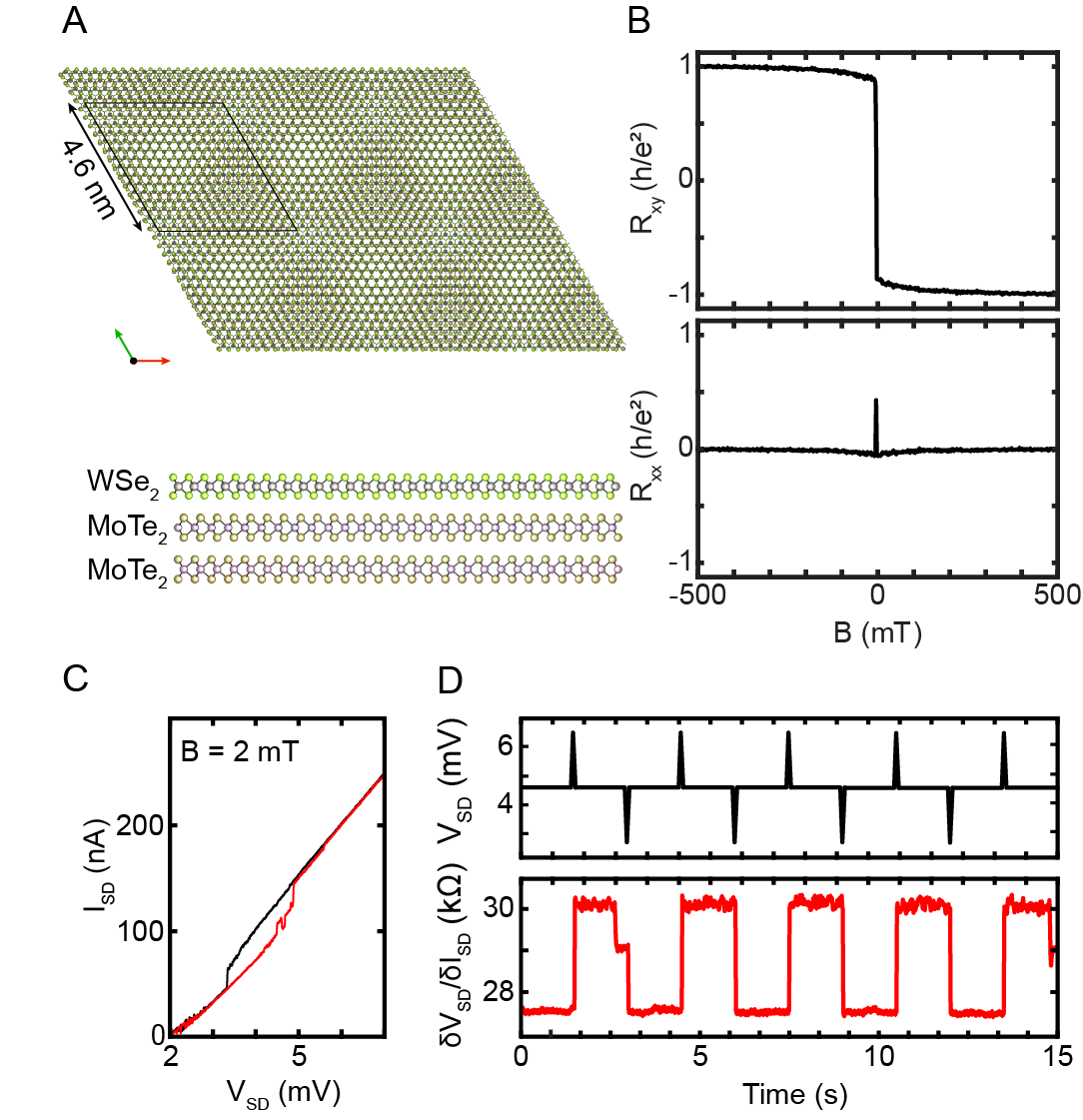}
\caption{
\textbf{Current-driven switching in AB-stacked 2L-MoTe$_2$/WSe$_2$ heterostructure.}
(\textbf{a}) Moir\'e superlattice structure for AB-stacked 2L-MoTe$_2$/WSe$_2$
(\textbf{b}) Electrical resistance measured at $T = 1.6$~K and moir\'e filling of $\nu=-1$. Hall resistance $R_{xy}$ is near the quantized value of $h/e^2$ while longitudinal resistance $R_{xx}$ vanishes, signatures of the QAH effect.
(\textbf{c}) Current induced resistance switching in the Chern magnet regime at 2 mT, measured in a two terminal configuration. Red and black curves correspond to rising and falling DC current. 
(\textbf{d}) Reproducible switching by DC current at 2 mT. Data in panels \textbf{b}, \textbf{c}, and \textbf{d} were obtained at $V_{TG} = -3.756$~V and $V_{BG} = 7.998$~V.}
\label{fig:fig1}
\end{figure}

\begin{figure*}[ht]
\includegraphics[width=7.25in]{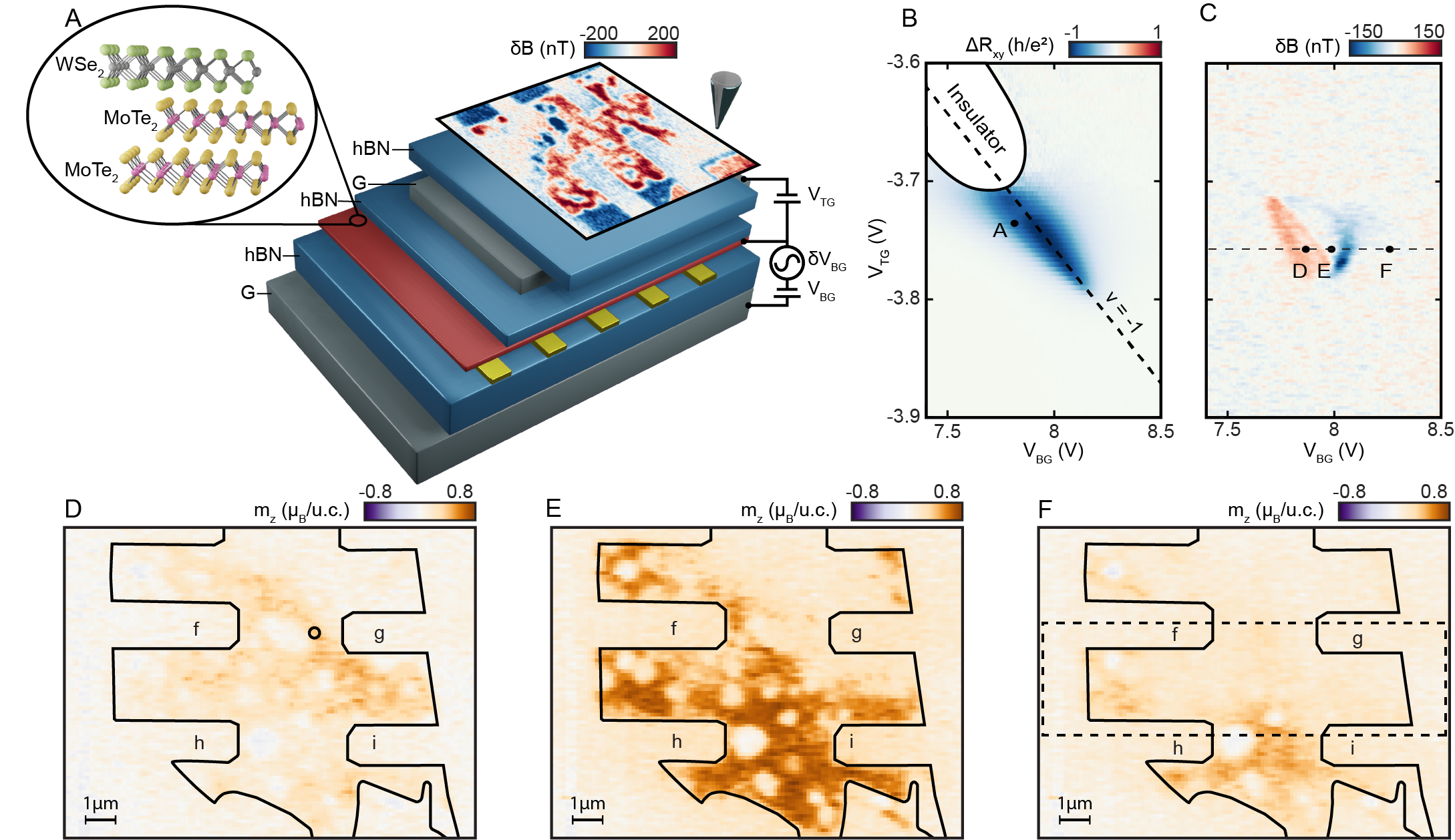}
\caption{
\textbf{Nanoscale magnetic imaging.}
(\textbf{a}) Device and measurement schematic. A finite frequency modulation $\delta V_{BG}$ is applied to the bottom gate, modulating the magnetic moment and producing a finite frequency magnetic field ($\delta B$) that penetrates the graphite top gate and is detected by a nanoSQUID sensor. 
(\textbf{b}) Anomalous Hall effect $\Delta R_{xy} \equiv \frac{1}{2}\cdot R_{xy}(B = 100$~mT $) - \frac{1}{2}\cdot R_{xy}(B = -100$~mT $)$ near $\nu=-1$ showing QAH phase.
(\textbf{c}) Phase diagram showing $\delta B$ as a function of $V_{TG}$ and $V_{BG}$ at a single point taken at $B = 38$~mT. 
(\textbf{d}-\textbf{f}) 2D maps of magnetization in region of device indicated in Fig. \ref{fig:sfigschematic}b. Images are acquired at voltages indicated by points in \textbf{c} at $B = 38$~mT. Circle in \textbf{d} indicates the position at which data in \textbf{c} was acquired. 
Quantization of $R_{xy}$ between contacts f,g,h,i occurs at parameters corresponding to panel \textbf{e}. A video showing magnetization as a function of $V_{BG}$ along the dotted line in \textbf{c} is available in the supplementary data.  
}
\label{fig:fig2}
\end{figure*}

Here, we study a trilayer heterostructure consisting of a MoTe$_2$ bilayer and WSe$_2$ monolayer stacked with 60$^{\circ}$ relative crystal axis alignment (see Methods and Fig. \ref{fig:sfigschematic}), producing a moir\'e pattern with wavelength $\lambda\approx 4.6$~nm (Fig. \ref{fig:fig1}a). 
When subjected to a strong perpendicular electric field, bands from the two semiconducting layers hybridize, producing topologically nontrivial moir\'e subbands\cite{zhang_spin-textured_2021}. At a moir\'e superlattice filling factor of $\nu=-1$, corresponding to one hole per superlattice unit cell, we observe a Chern magnet state characterized by a nearly quantized anomalous Hall effect and vanishingly small longitudinal resistance at $B=0$ (Fig. \ref{fig:fig1}b).
This is consistent with previous results on AB-stacked MoTe$_2$/WSe$_2$ bilayers\cite{li_quantum_2021}, where a magnetic Chern insulator arises due to the interplay of strong Coulomb repulsion and underlying Berry curvature of the moir\'e subbands\cite{pan_topological_2021, devakul_magic_2021, xie_valley-polarized_2022, devakul_quantum_2022,chang_theory_2022}. 

Measurements at finite current show hysteretic switching of the resistance in and near the Chern magnet state. Fig. \ref{fig:fig1}c shows the current measured as a function of rising and falling source-drain voltage bias at $\nu=-1$. We observe hysteretic switching of the current between at least two stable states. The switching current is approximately $100$~nA, corresponding to a current density of $j< 10^3$~A$\cdot$cm$^{-2}$. This is comparable to observations in twisted bilayer graphene\cite{sharpe_emergent_2019,serlin_intrinsic_2020} and significantly less than the lowest observed in spin-orbit torque devices\cite{jiang_efficient_2019,fan_magnetization_2014}. 
Switching is repeatable, as shown in Fig. \ref{fig:fig1}d.  

In order to investigate whether the current-driven metastability is related to magnetic domain dynamics, we image the magnetic structure in real space using a nanoscale superconducting quantum interference device (nanoSQUID)\cite{vasyukov_scanning_2013,anahory_squid--tip_2020}. The nanoSQUIDs are fabricated from indium\cite{anahory_squid--tip_2020} on the tip of a cryogenically cooled quartz tube, resulting in sensors with diameters ranging between $100-200$~nm and magnetic field sensitivities $\sim$15~nT/Hz$^{1/2}$. The quartz tube supporting the nanoSQUID is pressed against a piezoelectrically pumped quartz tuning fork, allowing the spatial position of the tip to be modulated in the plane of the sample, providing topographic feedback via shear-force microscopy. 

Figure \ref{fig:fig2}a shows a schematic of our measurement geometry.  Static voltages are applied to the top gate ($V_{TG}$) and bottom gate ($V_{BG}$) to control the charge carrier density and perpendicular electric displacement field on the grounded MoTe$_2$/WSe$_2$ trilayer (see Supplementary information)  In addition, a small AC voltage ($\delta V_{BG}$) is applied to the bottom gate at $f\approx 3$~kHz. Magnetic order, if present, is modulated by $\delta V_{BG}$, producing a change $\delta B$ in the fringe magnetic fields that may be detected by the nanoSQUID at $f_{AC}$.  
A real space map of $\delta B$ is shown in Fig. \ref{fig:fig2}a, acquired within the regime where the anomalous Hall resistance approaches quantization (see Fig. \ref{fig:fig2}b). Notably, magnetic fields penetrate the graphite top gate without significant modification\cite{brandt_magnetic_1988}, allowing us to explore the entire phase diagram tuned by $V_{TG}$ and $V_{BG}$. 
In addition, the high electronic compressibility of the graphite screens electrostatic potentials, preventing both unwanted local gating of the sample by the scanning tip and contamination of the magnetic signal by the weak but finite electric-field sensitivity of the nanoSQUID (see Fig. \ref{fig:sfigset}).   

Fig. \ref{fig:fig2}c shows $\delta B$ as a function of $V_{TG}$ and $V_{BG}$ measured at a single spatial coordinate. $\delta B$ is nonzero in the region of the phase diagram corresponding to the Chern magnet in transport measurements, and vanishes in regions for which $|\Delta R_{xy}| \ll h/e^2$.  
The magnetic field measured above a two dimensional layer is not, in general, a local probe of the magnetization $m_z$. To extract $m_z$---which we assume to be oriented in the out-of-plane direction---we first numerically integrate $\frac{\delta B_z}{\delta V_{BG}}(x,y)$ measured over a large real space area along a contour of constant $V_{TG}$ (shown in Fig. \ref{fig:fig2}c). This results in a map of the static $B_z(x,y)$, which can then be inverted through a Fourier domain magnetization inversion algorithm\cite{tschirhart_imaging_2021,thiel_probing_2019} to obtain $m_z(x,y,V_{BG})$ (see Fig. \ref{fig:sfigureacbgmagnetization}).
Images of $m_z(x,y)$ for several values of $V_{BG}$ are presented in Fig. \ref{fig:fig2}d-f, and a complete dataset depicting $m_z(x,y,V_{BG})$ is included in video format in the supplementary data. The active area of the moir\'e superlattice, defined by the intersection of the graphite top gate, the graphite bottom gate, the MoTe$_{2}$ bilayer, and the WSe$_{2}$ monolayer (see Fig. \ref{fig:sfigschematic}), is outlined in solid black line. 
We find a peak magnetization of $m_z\approx 1$~$\mu_B/u.c.$  The peak $m_z$ coincides with $\nu=-1$ and the QAH effect observed in transport.

The magnetization of the Chern magnet is spatially nonuniform. Throughout the QAH plateau, the bulk of the Chern magnet is riddled with submicron-sized holes. These holes do not become magnetized at any point in ($V_{TG},V_{BG}$) phase space (see Fig. \ref{fig:sfiguremagnetizationinhole}) and may correspond to local degradation of the air-sensitive MoTe$_2$ layer, decoupling of the moir\'e layers, or to the presence of competing structural allotropes of the MoTe$_2$. The presence of these defects does not seem to affect quantization of $R_{xy}$ in the QAH plateau, and the distribution of disorder is robust to thermal cycling (Fig. \ref{fig:sfigrepeatedcooldowns}).  
Inhomogeneity is also evident on much larger, $\sim 10$~$\mu m$ length scales. In particular, the maximal $m_z$ is achieved at different values of $V_{BG}$ in different parts of the device, consistent with long-range variations in the moir\'e unit cell area.  Microscopically, such variations may arise from interlayer strain or variations in the interlayer rotational alignment, though the latter are expected to play a smaller role in the physics of heterobilayers like MoTe$_2$/WSe$_2$ than in homobilayers such as twisted bilayer graphene\cite{lau_reproducibility_2022}.

\begin{figure*}[ht]
 \includegraphics[width=7.25in]{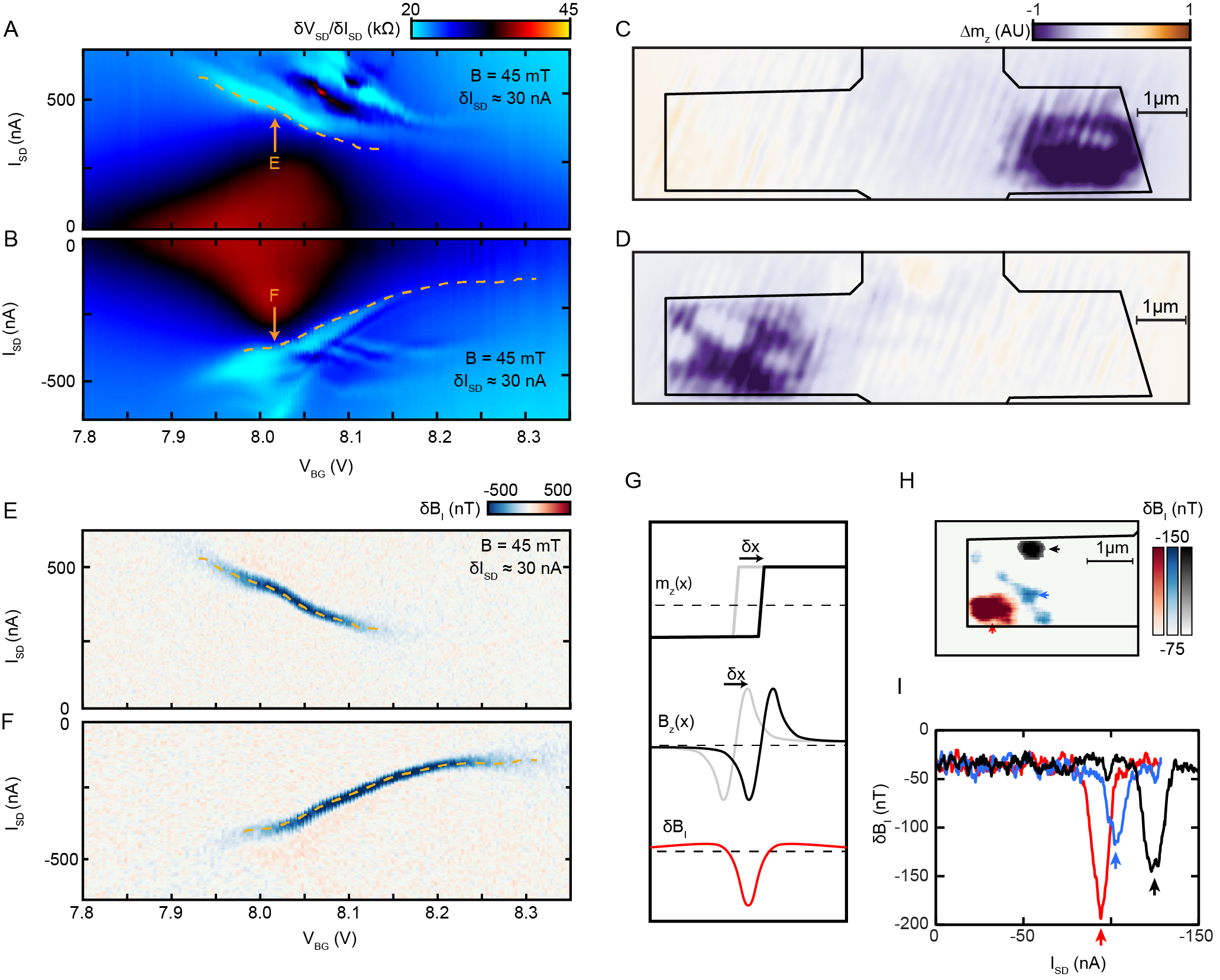}
\caption{
\textbf{Imaging current-induced switching.}
(\textbf{a}) Differential two-terminal resistance $\delta V_{SD}/ \delta I_{SD}$ as a function of $V_{BG}$ and $I_{SD}$ measured in the same configuration as data in Figs.\ref{fig:fig1}c-d. Quantization obtains near $V_{BG}\approx 8.0$~V. Sharp features in $\delta V_{SD}/ \delta I_{SD}$ appear at both finite bias and finite doping away from the QAH regime.  
(\textbf{b}) Differential two-terminal resistance $\delta V_{SD}/ \delta I_{SD}$ as a function of $V_{BG}$ and $I_{SD}$ with current sourced from the opposite direction (see Fig. \ref{fig:sfigure3explained}).  
(\textbf{c}) Current-induced change in magnetization $\Delta m_z=m_z(I_{SD})-m_z(I_{SD}=0)$ with $I_{SD} = 670$~nA, measured using AC gradient magnetometry (see Fig. \ref{fig:sfigureTFdomains}) and with $I_{SD} > 0$ corresponding to current from the bottom to the top of the scan range.  
(\textbf{d}) $\Delta m_z$ for $I_{SD} = -670$~nA. \textbf{c} and \textbf{d} were acquired in the QAH plateau.
(\textbf{e}) $\delta B_{I}$, defined as the local AC magnetic response to the AC current modulation $\delta I_{SD}$, measured near the left side of the device.
(\textbf{f}) The same, measured near the right side of the device. Local magnetic inversion corresponds closely to transport features in \textbf{a}. Gold dashed lines fit to the minima of $\delta B_{I}(V_{BG}, I_{SD})$ are overlaid on a, b, e, and f to illustrate this relationship precisely.    
(\textbf{g}) Origin of sharp peaks in $\delta B_{I}$. The AC current drives domain wall motion, generating a modulated magnetic field over the range of motion of the domain wall.  
(\textbf{h}) Spatial maps resolving domain wall motion. Red, blue, and black color scales indicate $108$, $119$, and $143$~nA $I_{SD}$, respectively. 
(\textbf{i}) Domain dynamics of current-stabilized magnetic domains as a function of current at positions indicated in \textbf{h}. \textbf{h} and \textbf{i} were acquired at $V_{TG} = -3.756$~V, $V_{BG} = 8.2$~V.
}
\label{fig:fig3}
\end{figure*} 

Equipped with a real space map of magnetic order, we may now investigate the origin of the current switching. 
Figures \ref{fig:fig3}a-b show a detailed dependence of the differential resistance, $\delta V_{SD}/\delta I_{SD}$, on $V_{BG}$ and $I_{SD}$. As in Fig. \ref{fig:fig1}c, current flows across the entire device, passing from the bottom to the top of the region depicted in Figs. \ref{fig:fig2}d-f.
The QAH plateau appears as a local maximum in $\delta V_{SD}/\delta I_{SD}$, and is centered around $V_{BG}\approx 8.0$~V. Features associated with current switching appear as sharp dips in differential resistance. 
Notably switching first occurs at values of $I_{SD}$ where quantization has already begun to degrade (see Fig. \ref{fig:breakdown}). 
We do not observe switching in the quantized regime where current flows only through the chiral edge states, suggesting that bulk current flow is required. 

\begin{figure*}[ht]
 \includegraphics[width=4.75in]{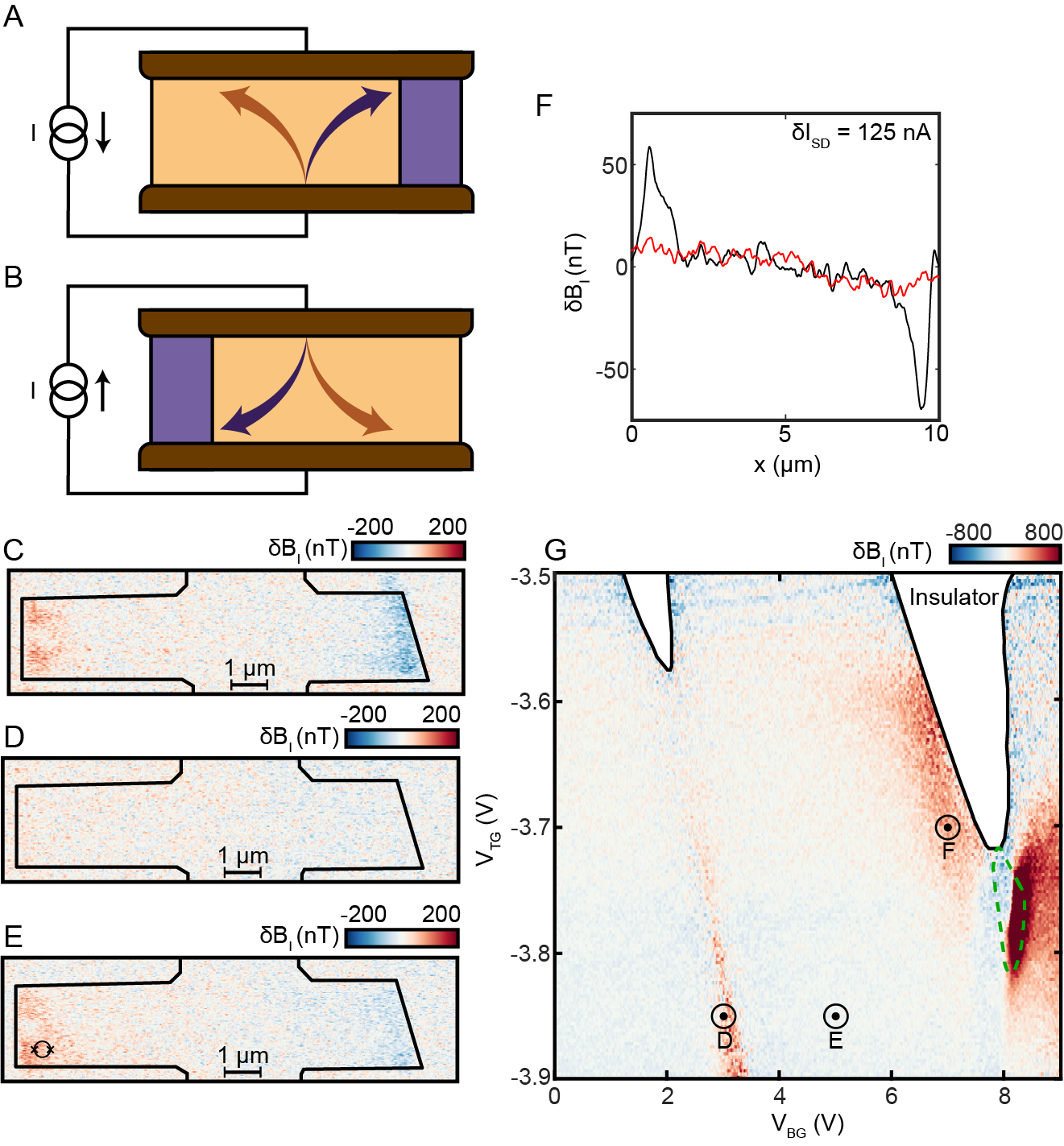}
 \caption{
 \textbf{Spin Hall effect}
(\textbf{a}-\textbf{b}) Schematic of intrinsic spin Hall torque mechanism.  When current enters the bulk, opposite spin states have opposite anomalous velocities, and thus accumulate on opposite sides of the device, exerting a spin orbit torque on the side that accumulates the minority spin state.  (\textbf{c}) Linecut across device with AC current at two different densities and displacement fields, showing spin Hall response ($\delta B$) in one regime and not the other.  (\textbf{d}) Scan of spin Hall effect near $\nu = 2$ (black linecut in (\textbf{a})).  (\textbf{e}) Absence of spin Hall effect far from commensurate filling (red linecut in (\textbf{a})).  (\textbf{f}) Scan of spin Hall effect near $\nu = 1$. (\textbf{g}) Phase diagram of spin Hall effect, taken at position indicated in \textbf{f}.  Locations in phase space of scans \textbf{d} - \textbf{f} are indicated, as is the presence of static magnetism (green dotted outline). Magnetic switching increases the signal dramatically when magnetism is present. 
}
\label{fig:fig4}
\end{figure*}

Current switching may be correlated precisely with magnetic structure.  
Figure \ref{fig:fig3}c shows the change in magnetization relative to the zero current state for $I_{SD}=670$~nA, well above the threshold current.  
The image is acquired using tuning fork based gradient magnetometry (see Fig. \ref{fig:sfigureTFdomains} and Methods) over the scan range depicted by the dashed box in Fig. \ref{fig:fig2}f.  
Above the threshold, a magnetic domain a few $\mu m^{2}$ in size is inverted relative to the ground state on one side of the device.  
Reversing the current flips the side hosting the reversed domain (Fig. \ref{fig:fig3}d). We conclude that the current switching corresponds to the reversal of magnetic domains, with the inverted domains appearing on opposite edges for opposing directions of applied DC current. This is confirmed by the fact that the required switching current increases dramatically as a function of the applied magnetic field (Fig. \ref{fig:sfigbdependence}), which increases the energy cost of an inverted magnetic domain.  

The correspondence between magnetic dynamics and resistivity may be probed in detail by examining the magnetic response, $\delta B_I$ to a small AC current.  Figs. \ref{fig:fig3}e and f show $\delta B_I$, measured near the right and left edges of the device, respectively, for the same range of $V_{BG}$, $V_{TG}$, and $V_{SD}$ as Fig. \ref{fig:fig3}a-b.  The local $\delta B_I$ signal shows a single sharp dip feature on the right side of the device for  $I_{SD}>0$ and on the left side for $I_{SD}<0$, but no signal for the opposite signs (see Fig. \ref{fig:sfigure3explained}).     
These features correlate precisely with the current switching features observed in transport, as evidenced by overlaying a fit to the local $\delta B_I$ dip on the transport data in Fig. \ref{fig:fig3}a-b. 

The $\delta B_I$ dips may be understood as a consequence of current-driven domain wall motion. As established above, applied current drives nucleation of minority magnetization domains. 
Once these domains are nucleated, increasing the current magnitude is expected to enlarge them through domain wall motion.  Where domain walls are weakly pinned, a small increase in the current $\delta I$ drives a correspondingly small motion $\delta x$ of the domain wall, producing a change in the local magnetic field $\delta B_I$ characterized by a sharp negative peak at the domain wall position (Fig. \ref{fig:fig3}g).  We may then use this mechanism to map out the microscopic evolution of domains with current. Fig. \ref{fig:fig3}h shows a spatial map of $\delta B_I$, measured at three different values of $I_{SD}$ corresponding to distinct features in the transport data (see Fig. \ref{fig:sfigbdependence}).  Evidently, the domain wall moves from its nucleation site on the device boundary towards the device bulk. Local measurements of $\delta B_I$ as a function of $I_{SD}$ show that this motion is itself characterized by threshold behavior, corresponding to the domain wall rapidly moving between stable pinning sites. A full correspondence of transport features and local domain dynamics is presented in Fig. \ref{fig:sfigureAllDomains}.

The symmetry of the observed magnetic switching is suggestive of a spin- or valley-Hall effect-driven mechanism\cite{shao_roadmap_2021}. 
In particular, magnetic inversion develops transverse to the applied current, such that the current-induced magnetization gradient, $\vec \nabla m_z \parallel \vec j \times \hat z$. In this mechanism, depicted schematically in Figs. \ref{fig:fig4}a-b, the current drives opposite spin or valley accumulation on the opposite sides of the device, consistent with our observations of magnetic inversion (Fig. \ref{fig:fig3}c-d).  
On the edge where the injected moments are not aligned with the equilibrium $m_z$, they may exert a torque on the ground state magnetic order, reversing it for sufficiently large steady state currents. We refer to  this mechanism as intrinsic spin Hall torque; it constitutes an intrinsic, single-layer version of conventional spin-orbit torque where a spin Hall effect layer is used to inject destabilizing moments into a second ferromagnetic layer \cite{shao_roadmap_2021}. The presence of this effect in MoTe$_2$/WSe$_2$ is not surprising, as the same Berry curvature that gives rise to the Chern magnet is expected to generate large spin- and valley- Hall effects, including at non-magnetic and non-integer band fillings. 

To investigate this hypothesis experimentally, we use local magnetic imaging to directly probe the current-driven accumulation of magnetic moments throughout the density- and displacement field-tuned phase space.  
Figs. \ref{fig:fig4}c-e show $\delta B_I$ maps measured at three different points, away from the regime where the ground state is ferromagnetic. A magnetic signal consistent with edge magnetic moment accumulation\cite{kato_observation_2004} is observed transverse to the applied current near both $\nu\approx-1$ and $\nu\approx-2$, though it is absent for $\nu\approx-1.6$. 
The accumulation of moments decays into the bulk with a length scale of several microns (Fig. \ref{fig:fig4}f), providing an estimate of the spin diffusion length consistent with measurements of the valley diffusion length in monolayer MoS$_2$\cite{lee_electrical_2016}. 
Measurements of $\delta B_I$ at a single point near the edge (Fig. \ref{fig:fig4}g) provide a reasonable proxy for the appearance of a spin- or valley-Hall effect and show that spin Hall-type signals, though ubiquitous, are concentrated in the vicinity of integer fillings $\nu=-1$ and $\nu=-2$. These fillings correspond to the Hubbard band edges, where the Berry curvature is expected to be enhanced by the appearance of correlation driven gaps, supporting an intrinsic origin for the spin Hall effect. 

We have shown here that the combination of intrinsic spin Hall effect with intrinsic magnetism provides a mechanism for a current-actuated magnetic switch in a single two dimensional electron system. 
The physical properties we invoke to explain this phenomenon are generic to all intrinsic Chern magnets.  
We emphasize that in both twisted bilayer graphene and our current MoTe$_2$/WSe$_2$ heterostructure, magnetic switching arises in regimes for which doping, elevated temperature, or disorder ensure that electrical current flows in the sample bulk. 
Ultra-low current switching of magnetic order has been observed in twisted bilayer graphene\cite{sharpe_emergent_2019,serlin_intrinsic_2020}; similar physics in that system is presumably governed by orbital, rather than spin, Hall effects\cite{he_giant_2020}. 
The bulk nature of the spin Hall torque mechanism means that similar phenomena should manifest not only in the growing class of intrinsic Chern magnets\cite{chen_tunable_2020,polshyn_electrical_2020,deng_quantum_2020}, but in all metals combining strong Berry curvature and broken time-reversal symmetry, including crystalline graphite multilayers\cite{zhou_half-_2021,zhou_isospin_2022}. 

Research into charge-to-spin current transduction has identified a set of specific issues restricting the efficiency of spin torque switching of magnetic order\cite{gupta_manipulation_2020,wang_current-driven_2019}.  
Spin current is not necessarily conserved, and as a result a wide variety of spin current sinks exist within typical spin torque devices. Extensive evidence indicates that in many spin torque systems a significant fraction of the spin current is destroyed or reflected at the spin-orbit material/magnet boundary\cite{schmidt_fundamental_2000}. 
In addition, the transition metals used as magnetic bits in traditional spin-orbit torque devices are electrically quite conductive, and can thus shunt current around the spin-orbit material, preventing it from generating spin current. 

These issues are entirely circumvented here through the use of a material that combines a spin Hall effect with magnetism, and as a result of these effects this spin Hall torque device has better current-switching efficiency than any known spin torque device.  

\section{Acknowledgements}
The authors acknowledge discussions with 
A. Macdonald, D. Ralph, Kelly Luo,  Vishakha Gupta, Rakshit Jain, Nai Chao Hu, Bowen Shen, and Zui Tao.
Work at UCSB was primarily supported by the Army Research Office under award W911NF-20-2-0166 and by the Gordon and Betty Moore Foundation EPIQS program under award GBMF9471. 
Work at Cornell was funded by the Air Force Office of Scientific Research under award no. FA9550-19-1-0390.
K.W. and T.T. acknowledge support from JSPS KAKENHI (Grant Numbers 19H05790, 20H00354 and 21H05233). 
ER and TA were supported by the National Science Foundation through Enabling Quantum Leap: Convergent Accelerated Discovery Foundries for Quantum Materials Science, Engineering and Information (Q-AMASE-i) award number DMR-1906325. 
CLT acknowledges support from the Hertz Foundation and from the National Science Foundation Graduate Research Fellowship Program under grant 1650114.


\section{Methods}

\subsection{Device fabrication}

AB-stacked 2L-MoTe$_2$/WSe$_2$ devices were fabricated using the layer-by-layer dry transfer method discussed in detail in \cite{li_quantum_2021}. An optical image of the device is presented in Fig. \ref{fig:sfigschematic}a. A dashed line identifies the extent of the few-layer graphene bottom gate. A black rectangle identifies a region illustrated in schematic form in Fig. \ref{fig:sfigschematic}b. Contact is made to the moir\'e superlattice formed by the MoTe$_2$ and WSe$_2$ crystals with $\approx5$~nm platinum contacts prepatterned onto a hBN flake; these are themselves contacted with gold wires outside the encapsulated region of the heterostructure.  The contacts used for the measurements presented here are labelled in Fig. \ref{fig:sfigschematic}b and are referred to throughout the main text using these labels.  Contacts f, g, h, and i are used to probe $R_{xy}$ and $R_{xx}$ in the Chern magnet. The relative locations of the WSe$_2$ monolayer, MoTe$_2$ bilayer, and few-layer graphene top gate are marked in red, blue, and light gray, respectively.  The region of overlap between these three flakes defines the device- the bottom gate is omitted for simplicity; it covers this region of overlap entirely, and thus does not define any edges of the dual-gated moir\'e superlattice.  A dashed rectangle identifies the region imaged using nanoSQUID magnetometry in Fig. \ref{fig:fig2}. The precise locations of the contacts relative to the device were determined using atomic force microscopy (AFM); this data is presented with an overlaid outline of the top gate (black line) and contacts (dashed lines) in \ref{fig:sfigschematic}c.  

Optical images of the WSe$_2$ monolayer, the few-layer graphene top gate, and the MoTe$_2$ bilayer are presented in Fig. \ref{fig:sfigschematic}d-f.  The crystal axes of the MoTe$_2$ and WSe$_2$ flakes were identified optically using angle-resolved second harmonic generation (SHG) and aligned with a 60$^{\circ}$ offset. In the case of the MoTe$_2$ bilayer (for which SHG cannot provide useful information) the crystal axes were identified for an attached monolayer (Fig. \ref{fig:sfigschematic}f). The relative positions of the top gate and WSe$_2$ flake were determined using optical microscopy during the stacking process (Fig. \ref{fig:sfigschematic}g).

\subsection{Electrical transport measurements}

The measurements presented here were conducted in a pumped liquid helium cryostat at a base temperature of 1.6K. 
AC transport data was acquired using a finite frequency excitation $\delta I\approx 300$~pA at $f\approx17$~Hz for Fig. \ref{fig:fig1}b and  \ref{fig:fig2}b and $f\approx3$~kHz for Fig. \ref{fig:fig3}a-b. 
Data in Figs. \ref{fig:fig1}b and \ref{fig:fig2} b are field symmetrized, so the plotted resistivity  R$_{xx}(B)$ = \((R_{meas}(B) + R_{meas}(-B))/2\) and  
R$_{xy}(B)$ = \((R_{meas}(B) - R_{meas}(-B))/2\). 

\subsection{Magnetic imaging}

Magnetic imaging is performed using a nanoscale superconducting quantum interference device (nanoSQUID). 
The typical static magnetic field in the region of space accessible by the nanoSQUID is below the DC noise floor of our sensor. Therefore, it is necessary to generate signals at finite frequency. We employ several different methods to generate the data presented in the main text and extended data figures including bottom gate modulation, spatial modulation of the nanoSQUID position, and modulation of the current in the device. Additional descriptions of these several techniques used to generate the data in the main text are available in the literature\cite{vasyukov_imaging_2017,anahory_squid--tip_2020,uri_nanoscale_2020,tschirhart_imaging_2021}.  

\subsubsection{Bottom gate modulation magnetometry}
As discussed in the main text, the magnetic signal may be modulated via carrier density variation, producing an AC response in the local magnetic field detected by the nanoSQUID. In practice, this is implemented using the circuit shown in Fig. \ref{fig:fig2}a. For the data shown in Figs. \ref{fig:fig2}a, c, d-e, Fig. \ref{fig:sfigureacbgmagnetization}, the bottom gate modulation has peak-peak amplitude of $\delta V_{BG}=35$~mV applied at frequency $f\approx 3$~kHz. We assume the resulting spatial map of $\delta B$ obeys $\delta B = \frac{dB_z}{dV_{BG}} \delta V_{BG}$.  

We reconstruct the magnetization under the assumption that it is entirely out-of-plane, so that $\vec m=m_z \hat z$. Reconstructed $m_z$ is shown in Figs. \ref{fig:fig2}d-e, \ref{fig:sfigureacbgmagnetization}, and \ref{fig:sfiguremagnetizationinhole}.  To do this, we first determine $B_z$, which is accomplished by acquiring $\delta B$ over a continuous range of $V_{BG}$ that spans the entire range of the magnetism.  As described in Fig. \ref{fig:sfigset}, the nanoSQUID is also sensitive to electric fields due to parasitic conduction through quantum dots near the tip. However, the $V_{BG}$ dependence of this signal is screened by the top gate, and varies slowly with $V_{BG}$ in the device regions outside the extent of the top gate. This spurious signal is eliminated by assuming that $B_z=0$ for values of $V_{BG}$ both lower and higher than the narrow domain of $V_{BG}$ where we observe magnetic structure in the bulk and in transport. Reconstruction of the magnetization from $B_z$ may then be done by Fourier transform techniques identical to those described in \cite{tschirhart_imaging_2021}. A schematic of this analysis is shown in Fig. \ref{fig:sfigureacbgmagnetization}, and video format data of the bottom gate evolution are available as supplementary data.   

\subsubsection{Spatial gradient magnetometry}
Figs. \ref{fig:fig3}c-d show a reconstruction of the steady state magnetization under an applied DC current. To avoid convolving modulations of the resistivity by the bottom gate with our detected SQUID signal, the magnetization is measured using gradient magnetometry. In this technique, we contact the nanoSQUID tip with a piezoelectric tuning fork (TF) which is modulated at $f\approx 32$~kHz. 
The resulting modulation of the in-plane nanoSQUID tip displacement produces a signal $\delta B_{TF} \approx \delta \vec r \cdot \vec \nabla_r B_z$, where  $\delta \vec r$ is the vector describing the spatial modulation of the tip. As described in both \cite{tschirhart_imaging_2021} and Fig. \ref{fig:sfigureTFdomains}, tuning fork measurements produce a number of additional spurious signals arising from electric fields and mechanical interactions with the surface. However, as in the bottom gate modulation magnetometry, these do not vary with the independent variable of interest here, the applied DC current $I_{SD}$. We thus analyze the difference images between zero and finite DC current, which contain only $\delta B_{TF}$. To convert $\delta B_{TF}$ to $B_z$, we integrate the signal along the direction of the oscillation, producing a map of $B_z$. This may then be converted to $m_z$ using the same standard Fourier domain analysis techniques described above and in \cite{tschirhart_imaging_2021}.  We did not precisely calibrate $\delta \vec r$ during this experimental run, and so provide the extracted $m_z$ in arbitrary units. However, prior work with the identical setup\cite{tschirhart_imaging_2021} allows us to estimate both the magnitude $|\delta \vec r|\approx 100$~nm and direction of $\delta \vec r$ (see Fig. \ref{fig:sfigureTFdomains}i).   

\subsubsection{Current modulation magnetometry}
As described in the main text, AC currents may also modulate magnetic structure and thus the local magnetic field signal. Data in Figs. \ref{fig:fig3}e,f,h,i, \ref{fig:fig4}c-g, \ref{fig:sfigbdependence}a,c,e-h, \ref{fig:sfigure3explained}e-f, \ref{fig:sfigureAllDomains}b-g, are all acquired in this way, with a current modulation $\delta I_{SD}$ applied at $f\approx3$~kHz. The contact configuration and amplitude of the applied current vary between data sets. Applying AC and DC bias to the source contact and grounding the drain, the parameters are: \begin{itemize}
\item Fig. \ref{fig:fig3}e, Fig. \ref{fig:sfigure3explained}f
        \subitem Source=j ; Drain=a,b,c,d
        \subitem $\delta I_{SD} = 30$~nA
\item Fig. \ref{fig:fig3}f, Fig. \ref{fig:sfigure3explained}c
        \subitem Source=a,b,c,d ; Drain=j
        \subitem $\delta I_{SD} = 30$~nA
\item Fig. \ref{fig:fig3}h-i, Fig. \ref{fig:sfigbdependence}e-h 
        \subitem Source=a,b,c,d ; Drain=j
        \subitem $\delta I_{SD} = 5$~nA
\item Fig. \ref{fig:sfigbdependence}a
        \subitem Source=a,b,c,d ; Drain=j
        \subitem $\delta I_{SD} = 10$~nA
\item Fig. \ref{fig:sfigbdependence}c
        \subitem Source=a,b,c,d ; Drain=j
        \subitem $\delta I_{SD} = 89$~nA
\item Fig. \ref{fig:fig4}c-f
        \subitem Source= j; Drain=a,b,c,d
        \subitem $\delta I_{SD} = 125$~nA
\item Fig. \ref{fig:fig4}g
        \subitem Source= j; Drain=a,b,c,d
        \subitem $\delta I_{SD} \in [0$ $270$~nA$]$
\item Fig. \ref{fig:sfigureAllDomains}b,d,f
        \subitem Source= a,b,c,d; Drain=j
        \subitem $\delta I_{SD}$ indicated on figure.
\item Fig. \ref{fig:sfigureAllDomains}b,d,f
        \subitem Source= j; Drain=a,b,c,d
        \subitem $\delta I_{SD}$ indicated on figure.
\end{itemize}
Throughout the main text figures, we standardize the phase of the AC current such that positive corresponds to increasing magnitude of current. 

\clearpage
\normalem
\let\oldaddcontentsline\addcontentsline
\renewcommand{\addcontentsline}[3]{}
\bibliography{references}

\clearpage
\widetext
\begin{center}
\end{center}
\renewcommand{\thefigure}{S\arabic{figure}}
\renewcommand{\thesubsection}{S\arabic{subsection}}
\setcounter{secnumdepth}{2}
\renewcommand{\theequation}{S\arabic{equation}}
\renewcommand{\thetable}{S\arabic{table}}
\setcounter{figure}{0}
\setcounter{equation}{0}
\onecolumngrid


\section{Extended data figures}

\begin{figure*}[ht]
\includegraphics[width=4.75in]{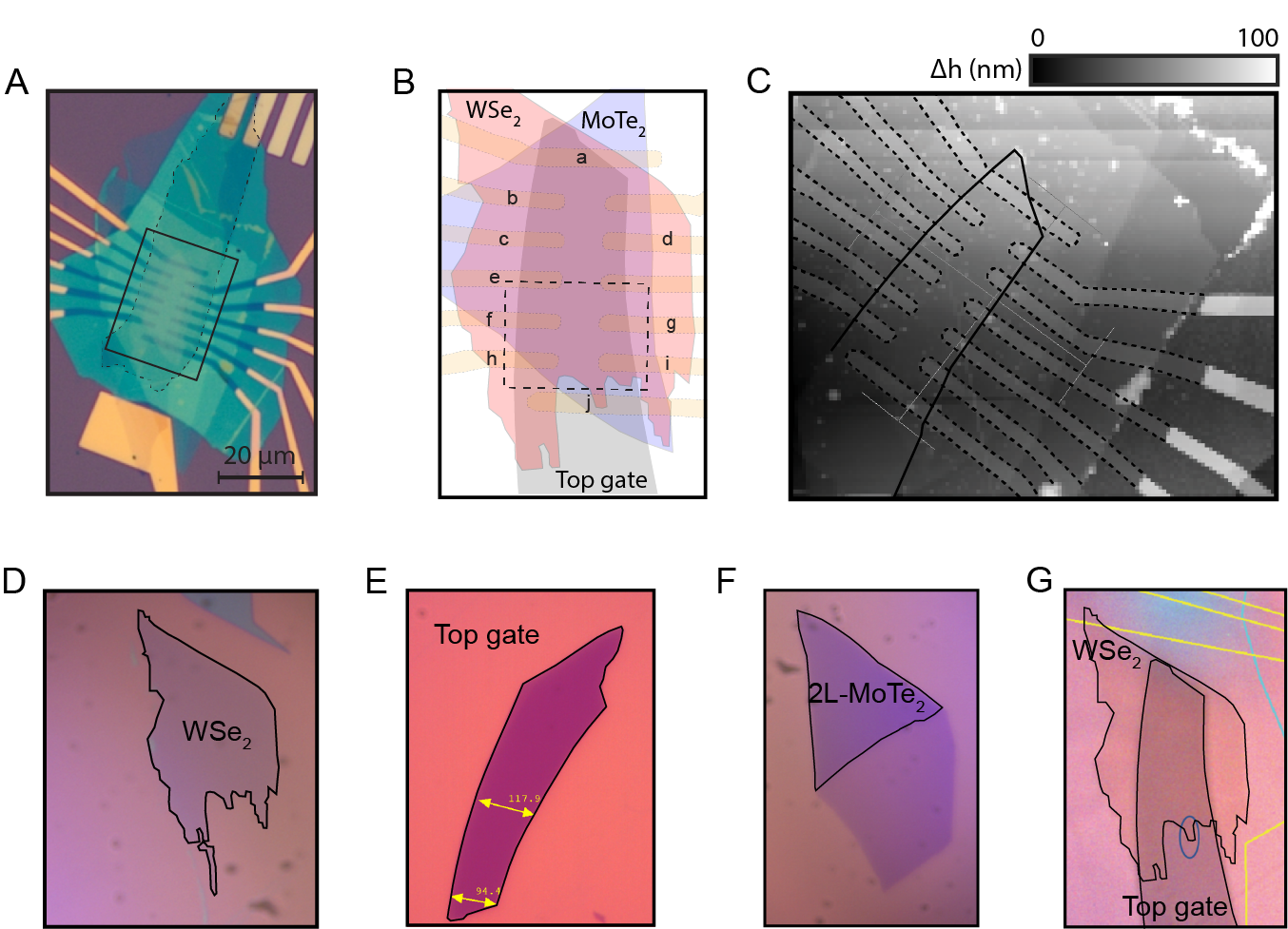}
\caption{
\textbf{Device schematic}
(\textbf{a}) Optical image of device.  Bottom gate is outlined with a dashed line.  
(\textbf{b}) Schematic of region outlined with a black line in (\textbf{a}).  Scan range shown in main text is outlined with dashed line.  
(\textbf{c}) AFM micrograph of device with contacts and top gate overlaid.  
(\textbf{d}) Optical image of WSe$_{2}$ monolayer.   
(\textbf{e}) Optical image of top gate.  
(\textbf{f}) Optical image of MoTe$_{2}$ bilayer attached to monolayer.  
(\textbf{g}) Optical image of top gate overlapping with WSe$_{2}$ monolayer.  
}
\label{fig:sfigschematic}
\end{figure*}

\begin{figure*}[ht]
 \includegraphics[width=4.25in]{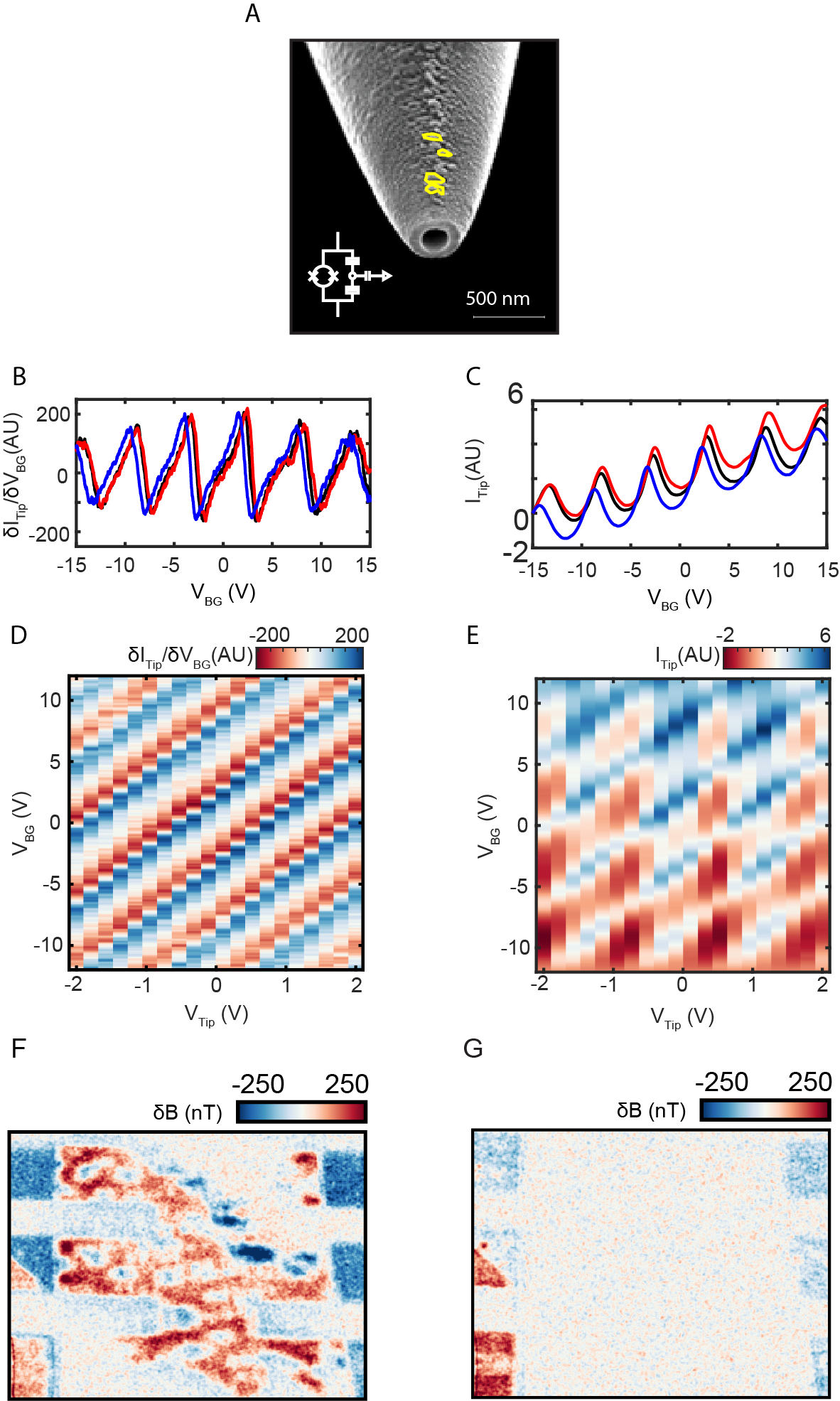}
\caption{
\textbf{Electric field sensitivity}
(\textbf{a}) Scanning electron micrograph of nanoSQUID tip with nanometer-scale metal droplets between the superconducting contacts outlined in yellow.  
Inset shows equivalent circuit, with SQUID in parallel with  Coulomb blockaded single electron transistor.  
(\textbf{b}) Differential conductance $\delta I_{Tip}/\delta V_{BG}$ of nanoSQUID tip as a function of heterostructure gate voltage $V_{BG}$ with nanoSQUID positioned above gate.  Blue, black, and red lines show three different measurements with different values of voltage applied to the tip ($V_{Tip})$.  
(\textbf{c}) Integrated current through nanoSQUID tip as a function of $V_{BG}$.  
(\textbf{d}) 2D plot of $\delta I_{Tip}/\delta V_{BG}$ as a function of both $V_{BG}$ and $V_{Tip}$
(\textbf{e}) 2D plot of $I_{Tip}$ as a function of both $V_{BG}$ and $V_{Tip}$, showing Coulomb blockade behavior of parasitic electric field sensitivity.  Such simple single-electron-transistor-like behavior is rare in nanoSQUIDs; electric field sensitivity is ubiquitous, but usually corresponds to a complex and disordered network of quantum dots and tunnel junctions.  
(\textbf{f}) NanoSQUID scan of device containing both magnetism and strong electric fields.  
(\textbf{g}) NanoSQUID scan of same region in a regime with no magnetism.  Note the lack of signal above the top gate and contacts, which screen out electric fields from the modulated bottom gate.  
}
\label{fig:sfigset}
\end{figure*}

\begin{figure*}[ht]
 \includegraphics[width=7.25in]{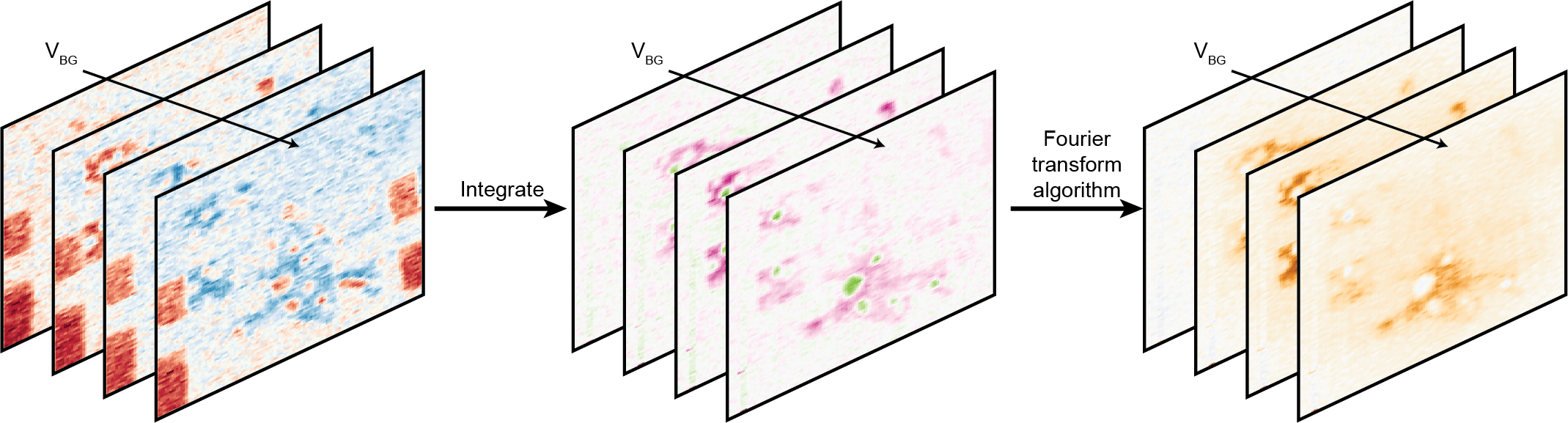}
\caption{
\textbf{AC bottom gate conversion to magnetization.}  
(\textbf{a}) AC bottom gate magnetometry produces $\delta B_{z}/\delta V_{BG} (x,y,V_{BG}) $.
(\textbf{b})  We integrate $\delta B_{z}/\delta V_{BG} (x,y,V_{BG})$ with respect to $V_{BG}$ to obtain $B_z(x,y,V_{BG})$
(\textbf{c})  For each value of $V_{BG}$ we invert $B(x,y)$ to produce $m_{z}(x,y,V_{BG})$, which is presented as a video of the out-of-plane magnetization as $V_{BG}$ enters and then leaves the QAH regime in the supplementary data.  
}
\label{fig:sfigureacbgmagnetization}
\end{figure*}

\begin{figure*}[ht]
 \includegraphics[width=4.75in]{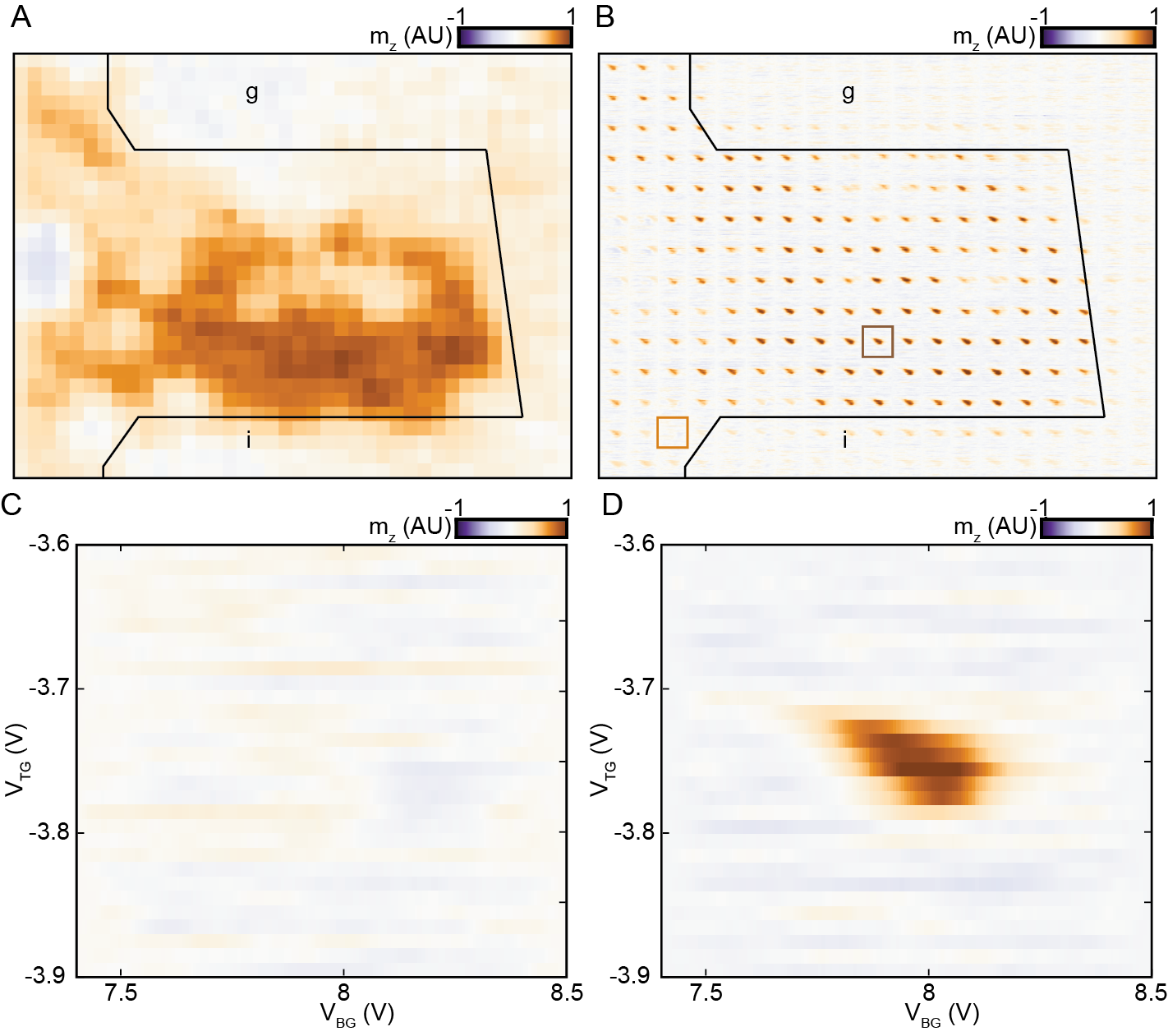}
\caption{
\textbf{Absence of magnetization in `holes.'}
(\textbf{a}) Magnetization of the device at $V_{TG} = -3.755$~V and $V_{BG} = 7.994$~V.
(\textbf{b}) Magnetization phase diagrams spatially mapped to the positions where they are taken. (\textbf{c}) Magnetization phase diagram in the `hole' highlighted by light-brown square in panel \textbf{b}.
(\textbf{d}) Magnetization phase diagram in the region highlighted with dark-brown on \textbf{b}. The signal in \textbf{c} is negligibly small compared to the signal in the magnetized region \textbf{d}, confirming the absence of the magnetized state in the `hole' for all values of applied gate voltages. 
}
\label{fig:sfiguremagnetizationinhole}
\end{figure*}

\begin{figure*}[ht]
 \includegraphics[width=3.5in]{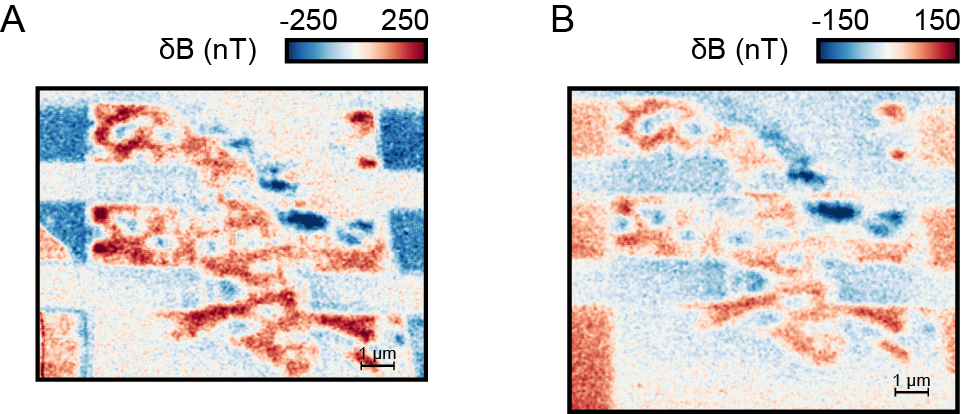}
\caption{
\textbf{Repeated cooldowns}
(\textbf{a}) Magnetic structure measured at $28$~mT, $V_{TG} = -3.734$~V, $V_{BG} = 7.818$~V at a height of 125 nm with nanoSQUID diameter $\phi = 159$ nm.
(\textbf{b}) Magnetic structure measured at $36$~mT, $V_{TG} = -3.756$~V, $V_{BG} = 7.943$~V during a subsequent cooldown at a height of 175 nm with nanoSQUID diameter $\phi = 113$ nm.
}
\label{fig:sfigrepeatedcooldowns}
\end{figure*}

\begin{figure*}[ht]
 \includegraphics[width=4.75in]{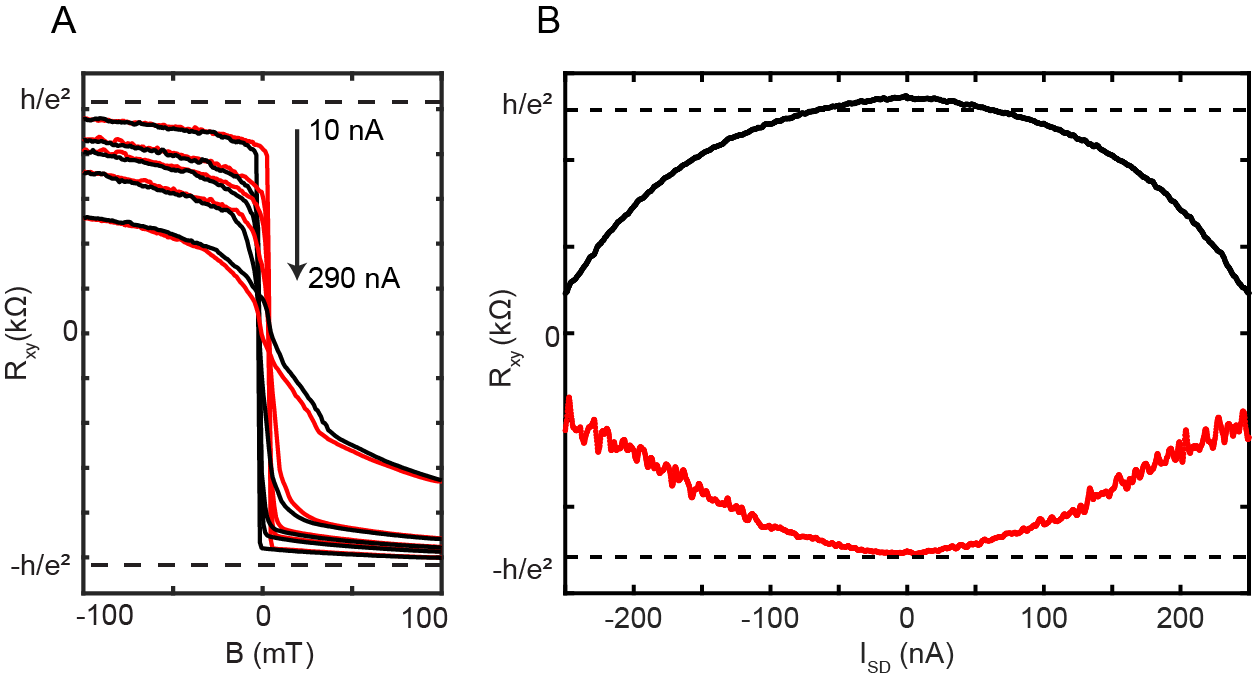}
\caption{
\textbf{Current-induced degradation of quantized transport.}
(\textbf{a}) $R_{xy}$ as a function of AC current. Red and black lines correspond to rising and falling magnetic field. Quantization of $R_{xy}$ is degraded by increased AC current.  
(\textbf{b}) Differential $R_{xy}$ measured using AC current and variable DC current $I_{SD}$. Red and black lines correspond to positive and negative magnetic fields, respectively.  
}
\label{fig:breakdown}
\end{figure*}

\begin{figure*}[ht]
 \includegraphics[width=4.75in]{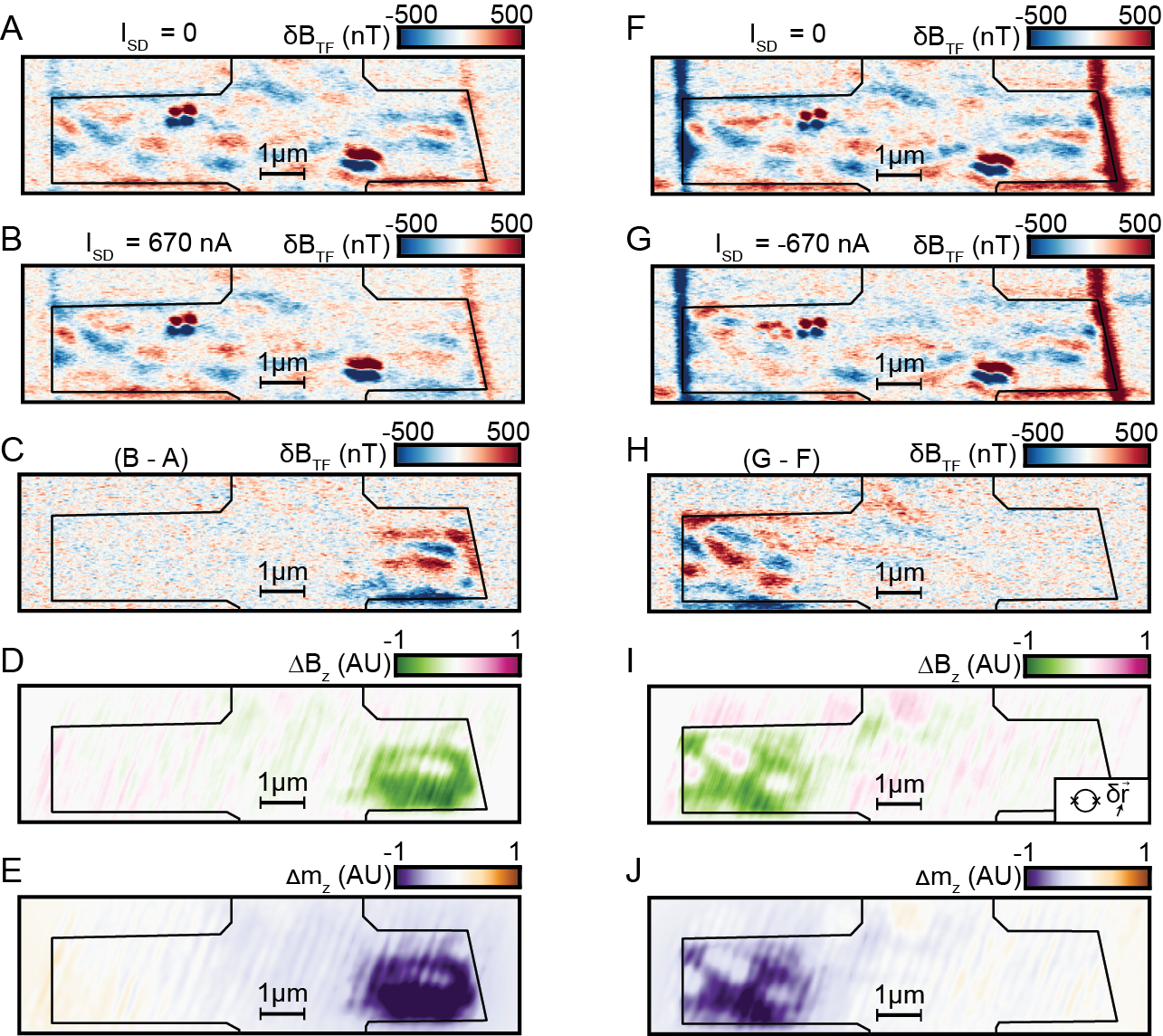}
\caption{
\textbf{Tuning fork imaging of static magnetization of domains.}  The two columns show the analysis pipeline used to generate Fig. \ref{fig:fig3}c (left column) and Fig. \ref{fig:fig3}d (right column).  
(\textbf{a}) Static background measured at $I = 0$~nA.  Signal includes both $\delta B_{TF}$ as well as electric field and mechanical signals.  
(\textbf{b}) $\delta B_{TF}$ measured at $I = +670$~nA. 
(\textbf{c}) Difference in $\delta B_{TF}$ obtained by subtracting data in panel \textbf{a} from data in panel \textbf{b}. (\textbf{d}) Magnetic field difference $\Delta B_z$ corresponding to fringe magnetic fields from the current-switched domain.  This is obtained by integrating $\delta B_{TF}$ with respect to in-plane coordinates along the vector $\delta \vec{r}$ indicated in the inset. Because the magnitude of $\delta \vec{r}$ is not known precisely, we present the data in arbitrary units.  
(\textbf e) Change in magnetization $\Delta m_z$ corresponding to the current switched domain for $I = +670$~nA. 
(\textbf{f-j}) The same as \textbf{a}-\textbf{e}, but for $I = -670$~nA. }
\label{fig:sfigureTFdomains}
\end{figure*}

\begin{figure*}[ht]
 \includegraphics[width=7.25in]{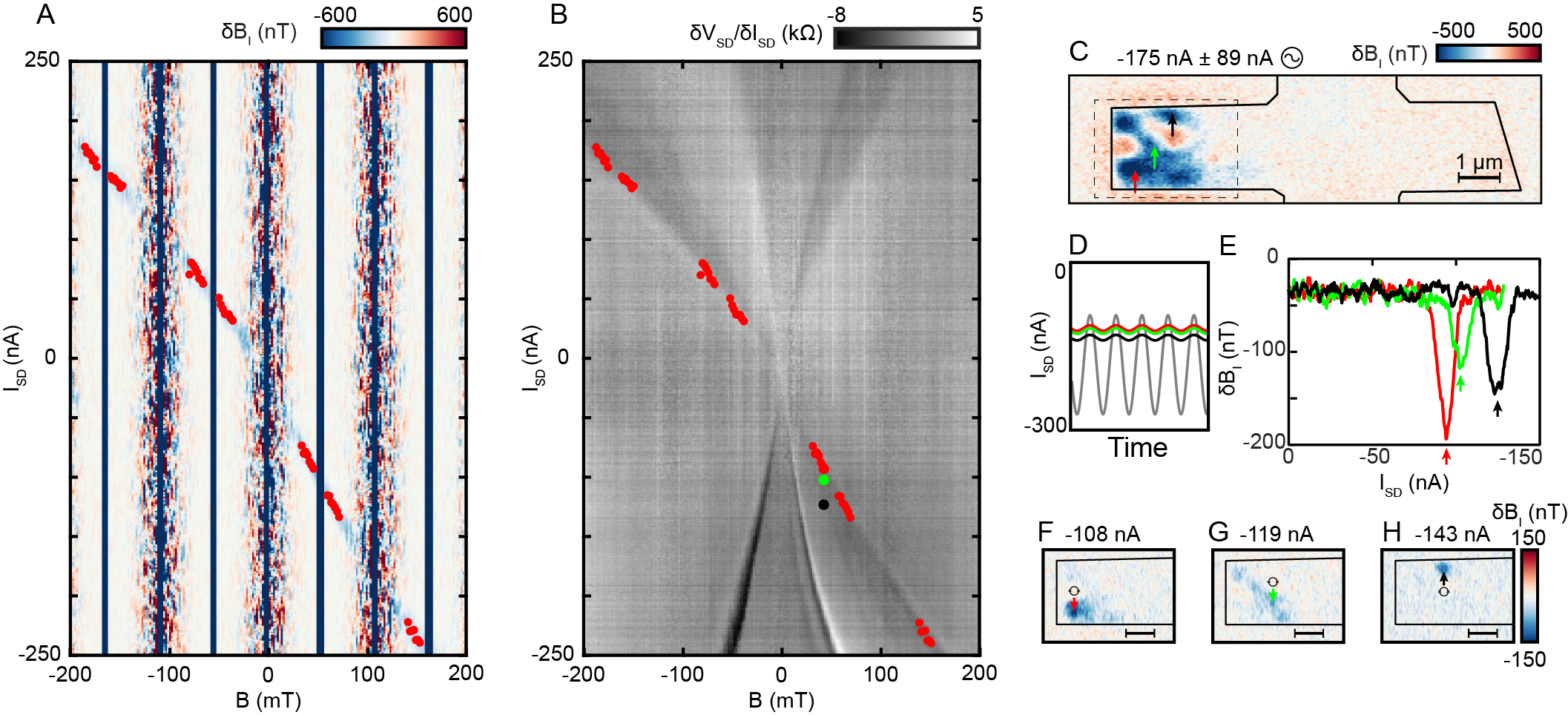}
\caption{
\textbf{Magnetic field dependence of domain dynamics}
(\textbf{a}) $\delta B_I$ measured for the domain on the left side of the device as a function of $I_{SD}$ and magnetic field at fixed $V_{BG} = 8.2$~V.
(\textbf{b})  Differential two-terminal resistance $\delta V_{SD}/ \delta I_{SD}$ as a function of $I_{SD}$ and $B$ measured simultaneously with data in panel \textbf{a}. $\delta V_{SD}/ \delta I_{SD}$ at 200 mT has been subtracted from all other values to highlight small variations.  Red markers identify minima of $\delta B_{I}$ from panel \textbf{a}.  Some variations in $\delta V_{SD}/ \delta I_{SD}$ coincide with and thus likely correspond to domain wall motion measured locally in panel \textbf{a}. Other variations may correspond to nucleation and pinning of magnetic domains walls at other locations.  
(\textbf{c}) $\delta B_I$ measurement with large $\delta I_{SD}=89$~nA, and fixed $I_{SD}=-175$~nA. 
(\textbf{d}) Illustration of the voltages applied for panels \textbf{c} (gray) and \textbf{f-h} (red, green, and black). The large AC voltage applied in panel \textbf{c} allows us to visualize the full range of domain wall positions for $-86$~nA$>I_{SD}>-264$~nA; smaller AC excitations applied in panels \textbf{e} (and \textbf{f}-\textbf{h}) allow us to resolve individual domain wall pinning sites within this range. 
(\textbf{e}) Dependence of $\delta B_{I}$ on $I_{SD}$ measured at three separate positions near the left edge of the device.  Domains are pinned at these positions at slightly different values of $I_{SD}$.
(\textbf{f-h}) Real space visualizations of the domain wall positions for the three peak values shown in panel \textbf{e}. The peaks correspond with weak features visible in transport data presented in panel \textbf{b}.   
}
\label{fig:sfigbdependence}
\end{figure*}

\begin{figure*}[ht]
\includegraphics[width=7.25in]{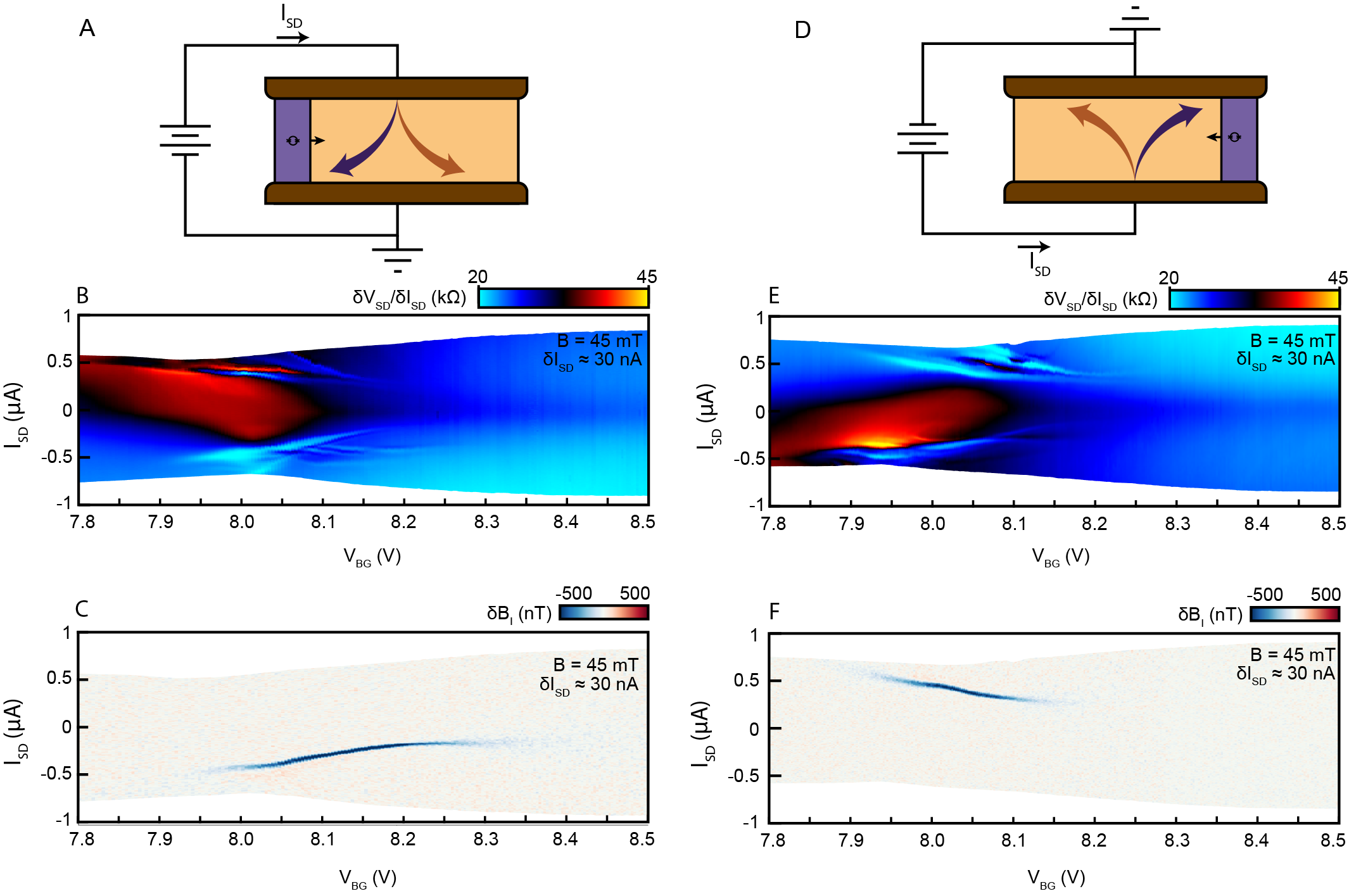}
\caption{
\textbf{Detail of current-driven domain dynamics I.}
(\textbf{a}) Schematic of measurement configuration 1, corresponding to current injected through the top contact with the bottom contact grounded. NanoSQUID is positioned near the left edge of the device.  
(\textbf{b}) Differential resistance $\delta V_{SD}/ \delta I_{SD}$ as a function of $I_{SD}$ and $V_{BG}$ for measurement configuration 1. 
(\textbf{c}) $\delta B_{I}$ measured on the left side of the device as a function of $I_{SD}$ and $V_{BG}$. This measurement was performed simultaneously with the measurement shown in \textbf{b}.   
(\textbf{d}) Measurement configuration 2, corresponding to current injected through the bottom contact with the top contact grounded. NanoSQUID is positioned near the right edge of the device.  The $I_{SD}$ DC sign convention is consistent between measurement configurations 1 and 2.  
(\textbf{e}) Differential resistance $\delta V_{SD}/ \delta I_{SD}$ as a function of $I_{SD}$ and $V_{BG}$ for measurement configuration 2.
(\textbf{f})  $\delta B_{I}$ measured on the right side of the device as a function of $I_{SD}$ and $V_{BG}$. This measurement was performed simultaneously with measurement shown in \textbf{e}.   
}
\label{fig:sfigure3explained}
\end{figure*}

\begin{figure*}[ht]
 \includegraphics[width=4.75in]{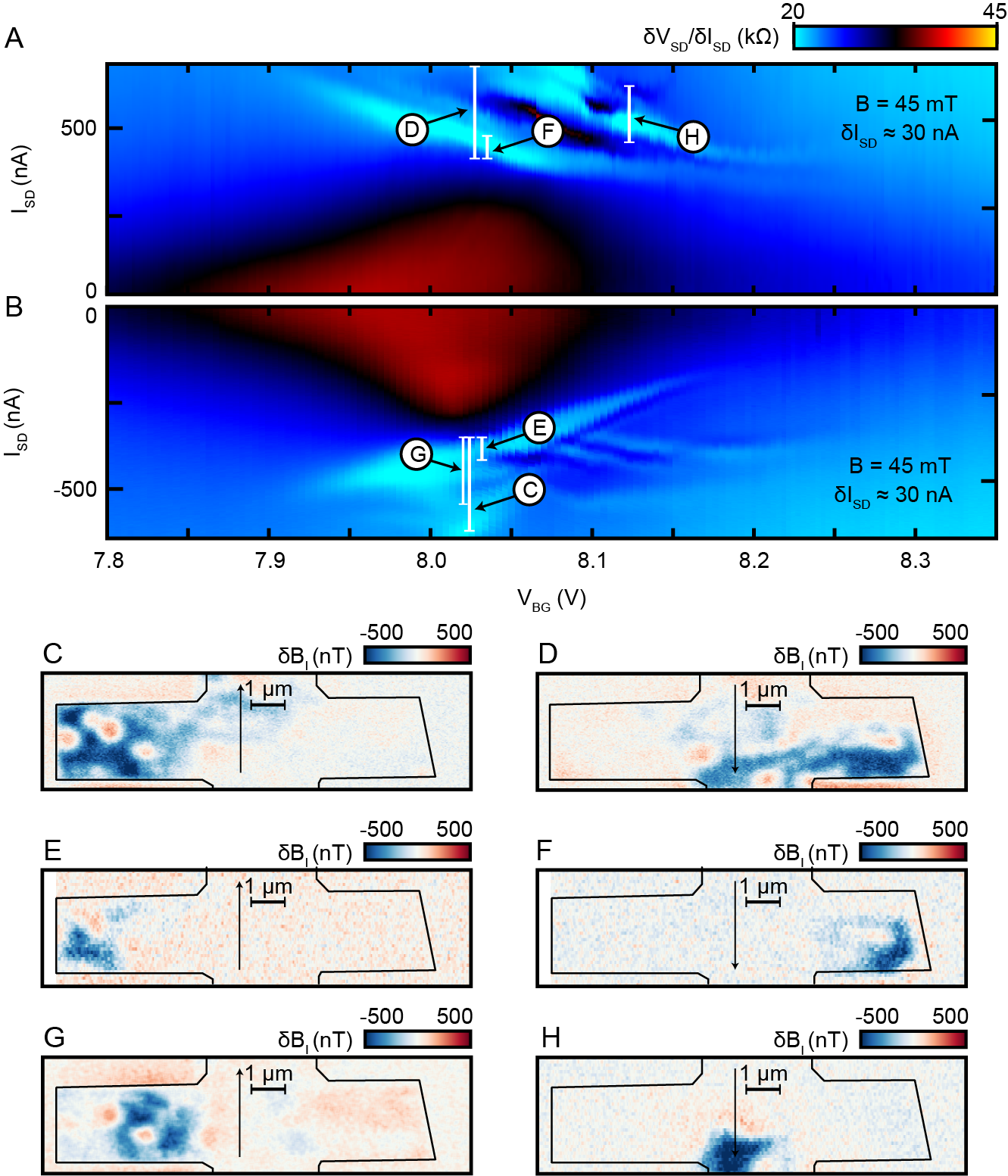}
\caption{
\textbf{Detail of current-driven domain dynamics II. }
(\textbf{a}) Differential two-terminal resistance $\delta V_{SD}/ \delta I_{SD}$. The data are the same as shown in Fig. \ref{fig:fig3}a and Fig. \ref{fig:sfigure3explained}b, $I_{SD}<0$. This dataset uses measurement configuration 1 from Fig. \ref{fig:sfigure3explained}a.  
(\textbf{b}) Differential two-terminal resistance $\delta V_{SD}/ \delta I_{SD}$. The data are the same as shown in Fig. \ref{fig:fig3}b. This dataset uses measurement configuration 2 from Fig. \ref{fig:sfigure3explained}b  and Fig. \ref{fig:sfigure3explained}e, $I_{SD}>0$.  
(\textbf{c}-\textbf{h}) 
$\delta B_{I}$ in response to AC current $\delta I_{SD}$ and DC current $I_{SD}$, for different values of $I_{SD}$ and $\delta I_{SD}$.  The intervals corresponding to the AC modulated current are indicated in panels \textbf{a} and \textbf{b}. 
}
\label{fig:sfigureAllDomains}
\end{figure*}

\clearpage
\section{Supplementary materials}

\begin{figure*}[ht]
 \includegraphics[width=4.75in]{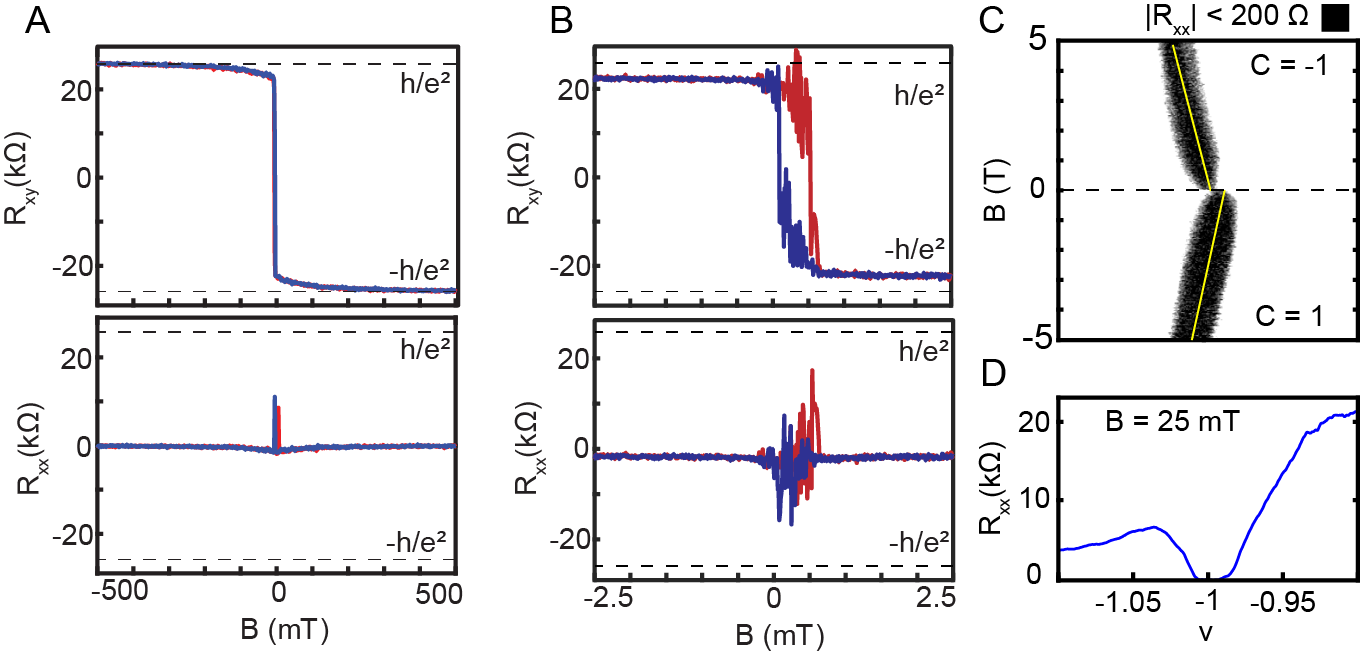}
\caption{
\textbf{Additional transport properties of the magnetic Chern insulator}
(\textbf{a}) Magnetic hysteresis loop in QAH regime.  At $\pm500$~mT quantization reaches $1.000\pm0.006$~$h/e^2$ and $-0.996\pm0.005$~$h/e^2$.
(\textbf{b}) Close to $B=0$ R$_{xy} \approx 0.9$~$h/e^2$.  Coercive fields are less than 1 mT at the measurement temperature of 1.6K.  
(\textbf{c}) Dependence of the degeneracy of the Chern band on B reveals the Chern number of the ground state at finite field, which is -1 in this system.  
(\textbf{d}) Linecut shows $R_{xx}$ reaches $-93$~$\Omega$ $\pm$ $115$~$\Omega$.  
}
\label{fig:sfig_additional_transport}
\end{figure*}

\begin{figure*}[ht]
\includegraphics[width=4.75in]{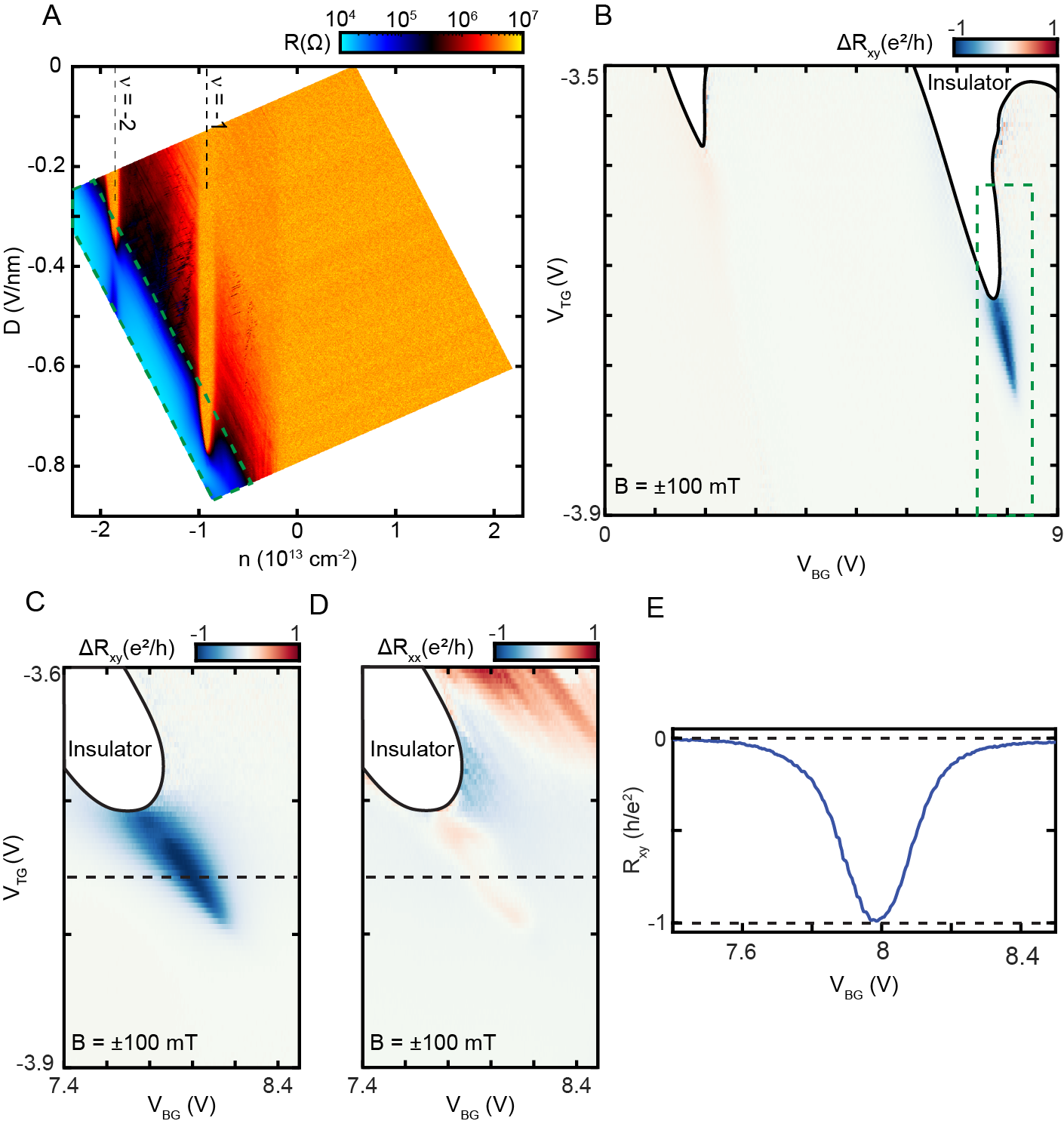}
\caption{
\textbf{Phase diagram from transport data}
(\textbf{a}) Two terminal resistance of device (measured between contacts $abcd$ and $j$) plotted as a function of electron density $n$ and displacement field $D$, where  \(D = \frac{V_{TG} - V_{BG}}{d_{t} + d_{b}} \), \(n = \epsilon \epsilon_0 (V_{TG}/d_{t} + V_{BG}/d_{b}) + n_0\). Here $d_{t}=2.7$~nm is the thickness of the top hBN layer, $d_b=12.1$~nm is the thickness of the bottom hBN layer, $\epsilon\approx3$ is the relative dielectric constant of hBN, and $n_0$ is the offset charge carrier density, $5.9\times10^{12}$~cm$^{-2}$.
At low displacement fields $\nu = -1$ and $\nu = -2$ both host topologically trivial interaction-driven insulating states.  In the electron-doping regime large contact resistances hamper transport measurements. 
(\textbf{b}) $\Delta R_{xy}$ measurement in the region inside the green dotted line in \textbf{a}.  A finite $\Delta R_{xy}$ appears near $\nu = -1$ in a narrow range of $D$. 
(\textbf{c}) $\Delta R_{xy}$ measurement in the region inside the green dotted line in \textbf{b}.  Precise quantization of $\Delta R_{xy}$ obtains over a range of displacement fields at $\nu = -1$.  
(\textbf{d}) Symmetrized $R_{xx}$ measurement in the region inside the dotted line in \textbf{b}.  
(\textbf{d}) Linecut of $\Delta R_{xy}$ along the dotted line in \textbf{c} illustrating the appearance of a QAH effect.}
\label{fig:sfiguretransport}
\end{figure*}

\begin{figure*}[ht]
 \includegraphics[width=4.75in]{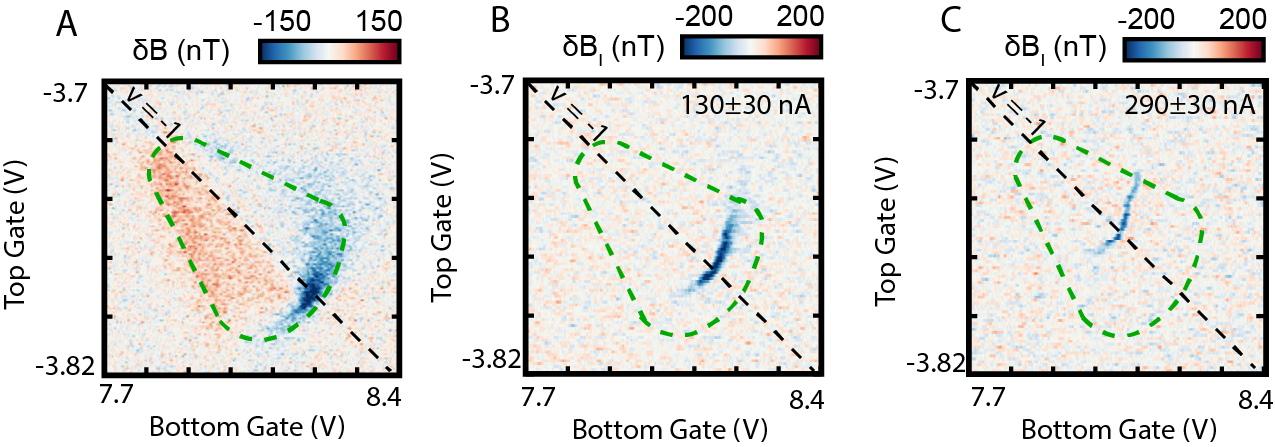}
\caption{
\textbf{Current-switching phase diagram}
(\textbf{a}) Bottom gate modulation magnetometry phase diagram taken at the point at which measurements for Fig. \ref{fig:fig4}g were performed. The green dotted outline appears in that figure.  $\delta V_{BG} = 35$~mV was used.  
(\textbf{b}) Current-induced magnetic domain switching signal $\delta B_{I}$ as a function of top and bottom gate voltages with $I_{SD} = 130$~nA,  $\delta I_{SD} = 30$~nA.
(\textbf{c}) The same measurement with $I_{SD} = 290$~nA, $\delta I_{SD} = 30$~nA.  Magnetic switching appears over broad regions of the magnetic phase diagram, although currents required to effect domain switching vary with gate voltages.  
}
\label{fig:sfigcurrentswitchingphasediagram}
\end{figure*}

\end{document}